\documentclass[11pt,graphicx]{article}
\usepackage{amssymb,amsmath,amsfonts}
\usepackage{graphicx}
\usepackage{graphics}
\usepackage{eepic,epsfig}
\newcommand{\n}{{\not\hspace{-0.5ex}\nabla}}
\newcommand{\A}{{\not\hspace{-0.8ex}A}}

\begin{document}
\title{Spinor Green function in higher-dimensional cosmic string space-time in the presence of magnetic flux}
\author{J. Spinelly$^{1}$ {\thanks{E-mail: jspinelly@uepb.edu.br}}  
and E. R. Bezerra de Mello$^{2}$ \thanks{E-mail: emello@fisica.ufpb.br}\\
1.Departamento de F\'{\i}sica-CCT, Universidade Estadual da Para\'{\i}ba\\
Juv\^encio Arruda S/N, C. Grande, PB, Brazil\\
2.Departamento de F\'{\i}sica-CCEN, Universidade Federal da Para\'{\i}ba\\
58.059-970, C. Postal 5.008, J. Pessoa, PB,  Brazil}

\maketitle
\begin{abstract}
In this paper we investigate the vacuum polarization effects associated with quantum fermionic charged fields in a generalized $(d+1)-$dimensional cosmic string space-times considering the presence of a magnetic flux along the string. In order to develop this analysis we calculate a general expression for the respective Green function, valid for several different values of $d$, which is expressed in terms of a bispinor associated with the square of the Dirac operator. Adopting this result, we explicitly calculate the renormalized vacuum expectation values of the energy-momentum tensors, $\langle T^A_B\rangle_{Ren.}$, associated with massless fields. Moreover, for specific values of the parameters which codify the cosmic string and the fractional part of the ratio of the magnetic flux by the quantum one, we were able to present in closed forms the bispinor and the respective Green function for massive fields. 
\vspace{1pc}
\end{abstract}
\maketitle
\section{Introduction}
Braneworld model has received renewed interest in recent years. By this scenario  our world is represented by a four dimensional sub-manifold, a three-brane, embedded in a higher dimensional spacetime \cite{Akama,Rubakov}. The idea that our Universe may have more than four dimensions was proposed by Kaluza \cite{Kaluza}, with the objective to unify  gauge theories with gravitation in a geometric formalism. Of particular interest are the models introduced by Randall and Sundrum \cite{RS,RS1}. The corresponding spacetime contains two (RSI), respectively one (RSII), Ricci-flat brane(s) embedded on a five-dimensional Anti-de Sitter (AdS) bulk. It is assumed that all matter fields are confined on the branes and gravity only propagates in the five dimensional bulk. In the RSI model, the hierarchy problem between the Planck scale and the electroweak one is solved if the distance between the two branes is about $37$ times the AdS radius.

Although topological defects have been first analysed in four-dimensional spacetime \cite{Kibble,V-S}, they have been considered in the context of braneworld. In this scenario the defects live in a $n-$dimensions submanifold embedded in a $(4+n)-$dimensional Universe. The domain wall case, with a single extra dimension, has been considered in \cite{Rubakov}. More recently the cosmic string case, with two additional extra dimensions, has been analysed \cite{Cohen,Ruth}. It has been shown that the gravitational effects of global strings can be responsible for compactification from six to four spacetime dimensions, naturally producing the observed hierarchy between electroweak and gravitational forces. 

Cosmic strings are topologically stable gravitational defects which appear in the framework of grand unified theories. These objects could be produced in very early Universe as a result of spontaneous breakdown of gauge symmetry \cite{Kibble,V-S}. Although the recent observation data on the cosmic microwave background have ruled out cosmic strings as the primary source for primordial density perturbation, they are still candidate for the generation of a number of interesting physical effects such as gamma ray burst \cite{Berezinski}, gravitational waves \cite{Damour} and high
energy cosmic ray \cite{Bhattacharjee}. Recently, cosmic strings have attracted renewed interest partly because a variant of their formation mechanism is proposed in the framework of brane inflation \cite{Sarangi}-\cite{Dvali}. 

The simplest theoretical model describing an idealized cosmic string, i.e., straight and infinitely thin, is given by a delta-type distribution for the energy-momentum tensor along the linear defect. As the solution of the Einstein equation, the geometry of the space-time produced by this source presents a conical singularity for the curvature tensor on its top. Under classical field theory viewpoint, this object can also be formed coupling the energy-momentum tensor associated with the Higgs $U(1)-$gauge system investigated by Nielsen and Olesen \cite{N-O} with the Einstein equations. This  project was successfully analysed by Garfinkle \cite{Garfinkle}. He found static cylindrically symmetric solutions representing vortices, as in flat space-time, and shown that asymptotically the space-time around the vortices is a Minkowski one minus a wedge. Their core have a non-zero thickness, and the magnetic fields vanishes outside them. Two years later Linet \cite{Linet} obtained, as a limit case, exact solutions for the metric tensor and Higgs field. He was able to show that the structure of the respective space-time corresponds to a conical one, with the conicity  parameter being expressed in terms of the energy per unity length of the vortex. 

The vacuum polarization effects associated with scalar and fermionic fields in the space-time of an idealized cosmic string, have been analyzed in \cite{scalar}-\cite{scalar5}, and \cite{ferm}-\cite{ferm3}, respectively. It has been shown that these effects depend on the parameter which codify the conical structure of the geometry.  Moreover, considering the presence of magnetic flux along the cosmic strings, there appears an additional contributions to the vacuum polarization effect associated with charged fields \cite{charged}-\cite{Spin}. Another type of vacuum polarization takes place when boundaries are presents. The imposed boundary conditions on quantum fields modifies the zero-point energy fluctuations spectrum and result in additional shifts in the vacuum expectation values of physical quantities, such as energy density and stress. This is the well-known Casimir effects.\footnote{For a review, see Ref. \cite{Most}} The analysis of Casimir effects in the idealized cosmic string space-time have been developed for a scalar \cite{Mello} and vector fields \cite{Mello1}, obeying boundary conditions on the cylindrical surfaces.\footnote{Also vacuum polarization effects induced by a composite topological defect has been analysed in \cite{Mello3}} 

The investigation of quantum effects in corresponding braneworld models is of considerable phenomenological interest, both in particle physics and cosmology. Quantum effects provide natural alternative for stabilizing the radion fields in a braneworld. The corresponding vacuum energy gives contribution to both the brane and bulk cosmological constant and, hence, has to be taking into account in the self-consistent formulation of the braneworld dynamics. Recently the fluxes by gauge fields play an important role in higher dimensional models including braneworld scenario (see for example \cite{Dou}). They provide stabilization mechanism for all moduli fields appearing, in particular, in various string compactifications. Motivated by these facts, we decided to analyse the vacuum polarization effects associated with a charged spin$-1/2$ field in a higher-dimensional cosmic string space-time considering an infinitely thin magnetic flux running along it. Specifically we are interested to calculate the vacuum expectation value (VEV) of the energy-momentum tensor. Although, as we shall see, for a $(d+1)-$dimensional cosmic string space-time, the renormalized VEV of the energy-momentum tensor associated with massless quantum fields presents the general form below
\begin{eqnarray}
	\langle T^A_B\rangle_{Ren.}=\frac{F^A_B}{r^{d+1}} \ ,
\end{eqnarray}
being $r$ the azimuthal distance to the string, the components of the tensor $F^A_B$ have different expressions depending on the dimension of the manifold. 

This paper is organized as follows: In Section \ref{sec2} we present a general expression for a charged massless spin$-1/2$ Green function valid for an arbitrary $(1+d)-$dimensional cosmic string space-time, $d\geq 2$. Although our main objective is to investigate the vacuum polarization effects in a six-dimensional spaces, the Green function obtained allows us to develop this analysis for lower dimensions as well. Moreover, in this section we also present a general expression for Green function associated with massive field for a particular choice of the parameters which codify the presence of cosmic string, and the fractional part of the ratio of the magnetic flux by the quantum one. In this way, the massive fermionic Green function is expressed in terms of a finite sum of modified Bessel function $K_\nu$, allowing us to present the renormalized VEV of the energy-momentum tensor in a closed form. In the Section \ref{sec3} we investigate the spinor Green functions in coincidence limit and extract all divergences in manifest form. Using this result, we provide explicit expressions for all components of the renormalized VEV of the energy-momentum tensor for different dimensions of the space, and for massless and massive fields. In Section \ref{conc}, we summarize the most important results obtained. In Appendix \ref{app} we present a generalization of the generation function for the modified Bessel functions, needed to construct the fermionic Green function as a finite sum of images of the Minkowski space-time functions. The Appendix \ref{app1} contains some technical formulas related with the calculation of the renormalized VEV of the energy-momentum tensor in Section \ref{sec3}.

\section {The spinor Green function}
\label{sec2}
This section is mainly devoted to calculate the Feynman propagator associated with a charged fermionic field propagating in a $(d+1)-$dimensional cosmic string space-time whose geometry is given by the line element
\begin{eqnarray}
\label{cs0}
	ds^2=g_{MN}dx^Mdx^N=-dt^2+dr^2+\alpha^2r^2d\varphi^2+\sum_{i}(dx^i)^2 \ . 
\end{eqnarray}
Specifically for a six-dimensional space, the coordinate system reads: $x^M=(t,r,\varphi,x,y,z)$, with $r\geq 0$, $\varphi\in[0, \ 2\pi]$, and $t, \ x^i\in(-\infty, \ \infty)$, $i=3, \ 4, \ 5$. $\alpha$ is parameter smaller than unity associated with the linear mass density of the string. In the braneworld scenario the space-time given by (\ref{cs0}), the bulk, represents a conical $3-$brane transverse to a two flat space.

In order to develop the calculation of the spinor Green function we shall adopt the $8\times 8$ Dirac matrices $\Gamma^M$ given below, which can be constructed in terms of the $4\times 4$ ones \cite{B-D,Moha}:
\begin{eqnarray}
\label{gamma}
\Gamma^0=\left( 
\begin{array}{cc}
0&\gamma^0 \\
\gamma^0&0 
\end{array} \right) \ , \
\Gamma^r=\left( 
\begin{array}{cc}
0&{\hat{r}}\cdot{\vec\gamma} \\
{\hat{r}}\cdot{\vec\gamma}&0 
\end{array} \right) \ , \
\Gamma^\varphi=\frac1{\alpha r}\left( 
\begin{array}{cc}
0&{\hat{\varphi}}\cdot{\vec\gamma} \\
{\hat{\varphi}}\cdot{\vec\gamma}&0 
\end{array} \right) \ , \nonumber\\
\Gamma^{x}=\left( 
\begin{array}{cc}
0&\gamma^{(3)}\\
\gamma^{(3)}&0 
\end{array} \right) \ , \
\Gamma^{y}=\left( 
\begin{array}{cc}
0&i\gamma_5\\
i\gamma_5&0 
\end{array} \right) \ , \
\Gamma^{z}=\left( 
\begin{array}{cc}
0&I \\
-I&0 
\end{array} \right) \ ,
\end{eqnarray}
where  $\gamma_5=i\gamma^0\gamma^1\gamma^2\gamma^3$, $I$ represents the $4\times 4$ identity matrix and ${\hat{r}}$ and ${\hat{\varphi}}$ stand the ordinary unit vectors in cylindrical coordinates. This set of matrices satisfies the Clifford algebra $\{\Gamma^M, \ \Gamma^N  \}=-2g^{MN}I_{(8)}$.

In the analysis of the vacuum polarization effects, we also consider the presence of an extra magnetic field running along the string. This magnetic field configuration is given by the following six-vector potential
\begin{eqnarray}
	A_M=A\partial_M\varphi
\end{eqnarray}
being $A=\frac{\Phi}{2\pi}$. 

The spinor Feynman propagator, defined as \cite{Birrel}
\begin{eqnarray}
	i{\cal{S}}_F(x,x')=\langle0|T(\Psi(x){\bar{\Psi}}(x'))|0\rangle \ ,
\end{eqnarray}
with ${\bar{\Psi}}=\Psi^\dagger\Gamma^0$, satisfies the non-homogeneous linear differential equation,
\begin{equation}
\label{Feyn}
\left(i\n +e\A -M\right){\cal{S}}_F(x,x')=\frac1{\sqrt{-g}}\delta^6(x-x')I_{(8)} \ ,
\end{equation}
where $g=det(g_{MN})$. The covariant derivative operator reads
\begin{equation}
\n=\Gamma^M(\partial_M+\Pi_M) \ ,
\end{equation}
being $\Pi_M$ the the spin connection given in terms of the $\Gamma-$matrices by
\begin{eqnarray}
\Pi_M=-\frac14\Gamma_N\nabla_M\Gamma^N \ ,
\end{eqnarray}
and
\begin{eqnarray}
\A=\Gamma^MA_M \ .
\end{eqnarray}
The Green function given in (\ref{Feyn}) is a bispinor, i.e., it transforms as $\Psi$ at $x$ and as ${\bar{\Psi}}$ at $x'$.

In \cite{Spin} we have shown that if a bispinor, ${\cal{D}}_F(x,x')$, satisfies the differential equation
\begin{eqnarray}
\label{D}
\left[{\Box}-ieg^{MN}(D_M A_N)+ie\Sigma^{MN}F_{MN}-2ieg^{MN}A_M\nabla_N\right. \nonumber\\
\left.-e^2g^{MN}A_M A_N-M^2-\frac14{\cal R}\right]{\cal{D}}_F(x,x')&=&-\frac1{\sqrt{-g}}\delta^6(x-x')I_{(8)} \ ,
\end{eqnarray}
with
\begin{equation}
\Sigma^{MN}=\frac14[\Gamma^M,\Gamma^N] \ , \ D_M=\nabla_M-ieA_M \ , 
\end{equation}
$\cal{R}$ being the scalar curvature and the generalized d'Alembertian operator 
given by
\begin{eqnarray}
\Box=g^{MN}\nabla_M\nabla_N=g^{MN}\left(\partial_M\nabla_N+\Pi_M\nabla_N-\{^S_{MN}\}\nabla_S\right) \ ,\nonumber
\end{eqnarray}
then the spinor Feynman propagator may be written as
\begin{equation}
\label{Sf} 
{\cal{S}}_F(x,x')=\left(i\n+e\A+M\right){\cal{D}}_F(x,x') \ .
\end{equation}

Applying this formalism for the system under investigation the operator $\cal{K}$, which acts on the left hand side of (\ref{D}), reads
\begin{eqnarray}
{\cal{K}}&=&{\Delta}+\frac i{\alpha^2 r^2}(1-\alpha)\Sigma^3_{(8)}\partial_\varphi-\frac1{4\alpha^2 r^2}(1-\alpha)^2+\frac e{\alpha^2 r^2} (1-\alpha)A\Sigma^3_{(8)}\nonumber\\
&-&\frac{2ie}{\alpha^2 r^2}A\partial_\varphi-\frac{e^2}{\alpha^2 r^2}A^2-M^2\ , 
\end{eqnarray}
where
\begin{equation}
\Sigma^3_{(8)}=\left( \begin{array}{cccc}
  \Sigma^3& 0\\ 
0 & \Sigma^3
                      \end{array}
               \right) \ , \ {\rm with} \
\Sigma^3=\left( \begin{array}{cccc}
  \sigma^3& 0\\ 
0 & \sigma^3
                      \end{array}
               \right) \ 
\end{equation}
and\footnote{For this geometry the only non-vanishing spin connection is $\Pi_\varphi=\frac i2(1-\alpha)\Sigma^3_{(8)}$.} 
\begin{equation}
\label{delta}
{{\Delta}}=-\partial_t^2+\partial_r^2+\frac 1r\partial_r+\frac 1{\alpha^2 r^2}\partial^2_\varphi+\partial_x^2+\partial_y^2+\partial_z^2 \ .
\end{equation}

Due to the fact of $\Sigma^3_{(8)}$ be a diagonal matrix, the bispinor ${\cal{D}}_F(x,x')$ is diagonal, too. In this way we can obtain an expression for this Green function analyzing only the effective $2\times 2$ matrix differential equation below:
\begin{eqnarray}
\label{K}
\left[{\Delta}+\frac i{\alpha^2 r^2}(1-\alpha)\sigma^3\partial_\varphi-\frac1{4\alpha^2 r^2}(1-\alpha)^2+\frac e{\alpha^2 r^2} (1-\alpha)A\sigma^3\right.\nonumber\\
\left.-\frac{2ie}{\alpha^2 r^2}A\partial_\varphi-\frac{e^2}{\alpha^2 r^2}A^2-M^2\right]{\cal{D}}^{(2)}_F(x,x')= -\frac1{\sqrt{-g}}\delta^6(x-x')I_{(2)} \ .
\end{eqnarray}
So the complete Green function is given in terms of ${\cal{D}}^{(2)}_F(x,x')$, in a diagonal matrix form.

On basis of these results, for this six-dimensional cosmic string space-time, the fermionic propagator reads:
\begin{eqnarray}
	S_F(x,x')=\left[i\Gamma^0\partial_t+i\Gamma^r\partial_r+i\Gamma^\varphi\partial_\varphi+i\Gamma^i\partial_i-\frac{1-\alpha}2\Gamma^\varphi\Sigma^3_{(8)}+\frac {e\Phi}{2\pi}\Gamma^\varphi+M\right]D_F(x,x') \ .
\end{eqnarray}
                     
The vacuum average value for the energy-momentum tensor can be expressed in terms of the Euclidean Green function. It is related with the ordinary Feynman Green function \cite{Birrel} by the relation  ${\cal D}_E(\tau,\vec{r}; \tau', \vec{r'}) = -i {\cal D}_F(x,x')$, where $t=i\tau$. In the following we shall consider the Euclidean Green function. 

In order to obtain the Euclidean Green function ${\cal{D}}^{(2)}_E(x,x')$ in explicit form, let us find the complete set of bispinor which obey the eigenvalue equation
\begin{eqnarray}
\label{K1}
	{\bar{\cal{K}}}\Phi_\lambda(x)=-\lambda^2\Phi_\lambda(x)
\end{eqnarray}
with $\lambda^2\geq 0$ and being ${\bar{\cal{K}}}$ the Euclidean version of the differential operator given in (\ref{K}). So we may write
\begin{eqnarray}
\label{D2}
	{\cal{D}}^{(2)}_E(x,x')=\sum_{\lambda^2}\frac{\Phi_\lambda(x)\Phi_\lambda^\dagger(x')}{\lambda^2}=\int_0^\infty \ ds\sum_{\lambda^2} \Phi_\lambda(x)\Phi_\lambda^\dagger(x') \ e^{-s\lambda^2} \ .
\end{eqnarray}
The eigenfunctions of (\ref{K1}) can be specified by a set of quantum number associated with operators that commute with ${\bar{\cal{K}}}$ and among themselves: $p_\tau=-i\partial_\tau$, $p_i=-i\partial_i$ for $i=3,\ 4, 5$ , $L_\varphi=-i\partial_\varphi$ and the spin operator, $\sigma_3$. Let us denote these quantum numbers by $(k^\tau, \ k^i, \ n, \ \sigma)$, where $(k^\tau, \ k^i)\in (-\infty, \ \infty)$, $n=0, \ \pm 1, \ \pm 2, \ ... \ $, $\sigma=\pm 1$. Moreover, these functions also depend on the number $p$, which satisfies the relation $\lambda^2=p^2+k^2+M^2$ and assumes values in the interval $[0, \ \infty)$. 

Although we have developed this formalism for a six-dimensional cosmic string space-time, it can be adapted for a three, four and five dimensional spaces. The reason resides in the representations adopted for the Dirac matrices in these dimensions. For three dimensional case the Dirac matrices are given in terms of $2\times 2$ Pauli matrices. For four and five dimensions, the Dirac matrices are given by the $4\times 4$ off-diagonal matrices from the first four and five matrices given in (\ref{gamma}). Of course for these cases we have to discard the derivative associated with the respective extra coordinates in (\ref{delta}). So on basis on these arguments it is possible to generalize the eigenfunction of the operator (\ref{K1}) by:

\begin{eqnarray}
\label{Phi}
	\Phi_\lambda^{(+)}(x)&=&\frac{e^{ik.x}{\sqrt{p}}}{[\alpha(2\pi)^{N+1}]^{1/2}}e^{in\varphi}J_{|\nu^+|/\alpha}(pr){\cal{\omega}}^{(+)} \ ,  \nonumber\\
	\Phi_\lambda^{(-)}(x)&=&\frac{e^{ik.x}{\sqrt{p}}}{[\alpha(2\pi)^{N+1}]^{1/2}}e^{in\varphi}J_{|\nu^-|/\alpha}(pr){\cal{\omega}}^{(-)}
\end{eqnarray}
where
	\begin{eqnarray}
	\label{omega}
	{\cal{\omega}}^{(+)}=\left(
\begin{array}{cc}
1 \\
0
\end{array} \right) \  , \ 
	{\cal{\omega}}^{(-)}=\left(
\begin{array}{cc}
0 \\
1
\end{array} \right) \ 
\end{eqnarray}
are the eigenfunctions of the the operator $\sigma_3$, $J_\mu(z)$ is the Bessel function, and $\nu^{\pm}=n\pm\frac{(1-\alpha)}2-({\bar{N}}+\gamma)$, where we have defined $eA=e\frac{\Phi}{2\pi}={\bar{N}}+\gamma$, the ratio of the magnetic flux by the quantum one, in terms of an integer number, ${\bar{N}}$, and its fractional part, $\gamma$. In (\ref{Phi}) we may assume $N=1, \ 2, \ 3, \ 4$, which are related with the dimensions of the space considered: three, four, five or six, as explained before.

Now we are in position to calculate the Green function by using (\ref{D2}), (\ref{Phi}) and (\ref{omega}) as show below
\begin{eqnarray}
	\label{D2a}
		{\cal{D}}^{(2)}_E(x,x')&=&\frac1{\alpha(2\pi)^{N+1}}\int_0^\infty ds  \int d^Nk \ \int_0^\infty \ dp \ p  \ e^{ik(x-x')}\sum_n e^{in(\varphi-\varphi')}\nonumber\\
	&& {\rm diag} (J_{|\nu^+|/\alpha}(pr)J_{|\nu^+|/\alpha}(pr'), J_{|\nu^-|/\alpha}(pr)J_{|\nu^-|/\alpha}(pr')) \ e^{-s(p^2+k^2+M^2)} \ .
\end{eqnarray}
With the help of \cite{Grad} we can express (\ref{D2a}) by
\begin{eqnarray}
\label{D2b}
	{\cal{D}}^{(2)}_E(x,x')&=&\frac1{\alpha(4\pi)^{N/2+1}}\int_0^\infty \frac{ds}{s^{N/2+1}}e^{-\frac{(\Delta x)^2+r^2+r'^2}{4s}-M^2s} \ \sum_n e^{in(\varphi-\varphi')}\nonumber\\
	&&{\rm diag}(I_{|\nu^+|/\alpha}(rr'/2s), \ I_{|\nu^-|/\alpha}(rr'/2s)) \ ,
\end{eqnarray}
where $I_\mu(z)$ is the modified Bessel function.

The calculations of the renormalized VEV of the energy-momentum tensor associated with massive quantum fields in cosmic string space-time, have been developed by many authors cited in \cite{ferm,charged}. All of these calculations involve very complicated integrals. In order to present more suitable expressions, the authors make applications taking the massless limit in previous results. Here, in this paper, we shall explicitly exhibit the renormalized vacuum expectation value of the fermionic energy-momentum tensor for different dimensions of the space-time in closed expressions. So, in order to do that we shall consider two different limiting cases which allow us to reach this objective as shown in the following subsections. 

\subsection{Massless fermionic fields}
In this subsection we shall calculate the bispinor associated with a charged massless fermionic fields in an arbitrary dimensional cosmic string space-time. So taking $M=0$ in (\ref{D2b}) we obtain \cite{Grad}
\begin{eqnarray}
\label{D2c}
	{\cal{D}}^{(2)}_E(x,x')&=&\frac1{\alpha(2\pi)^{\frac{N+3}2}}\frac1{(rr')^{\frac{N}2}}\frac1{(-\sinh u)^{\frac{N-1}2}} \times \nonumber\\
&&\sum_n e^{in(\varphi-\varphi')}\left( 
\begin{array}{cc}
Q_{|\nu^+|/\alpha-1/2}^{\frac{N-1}2}(\cosh u)&0\\
0& Q_{|\nu^-|/\alpha-1/2}^{\frac{N-1}2}(\cosh u)) 
\end{array} \right)  \ \ ,
\end{eqnarray}
where
\begin{eqnarray}
\label{cs}
	\cosh u=\frac{(\Delta x)^2+r^2+r'^2}{2rr'} \ .
\end{eqnarray}
In the above expression, $Q^\lambda_\nu$ is the associated Legendre function. It is possible to express this function in terms of hypergeometric function \cite{Grad} by:
\begin{eqnarray}
Q^\lambda_\nu(\cosh u)&=&e^{i\lambda\pi}2^\lambda{\sqrt{\pi}} \ \frac{\Gamma(\nu+\lambda+1)}{\Gamma(\nu+3/2)} \frac{e^{-(\nu+\lambda+1)u}}{(1-e^{-2u})^{\lambda+1/2}}(\sinh u)^\lambda\times \nonumber\\
&&F\left(\lambda+1/2 \ , \ -\lambda+1/2 \ ; \ \nu+3/2 \ ; \ \frac1{1-e^{2u}}\right) \ .
\end{eqnarray}
In this analysis we identify the parameter $\nu$ with $|\nu^\pm|/\alpha-1/2$ and take $\lambda=(N-1)/2$. So the relevant hypergeometric function is
\begin{eqnarray}
	F\left(\frac N2 \ , \ -\frac{N-2}2 \ ; \ |\nu^\pm|/\alpha+1 \ ; \ \frac1{1-e^{2u}}\right) \nonumber \ .
\end{eqnarray}
If $N$ is a even number, this function becomes a polynomial of degree $\frac{N-2}2$; however being $N$ an odd number, this function is an infinite series.

As special applications of this formalism, let us consider specific values for $N$ as shown in the following.

\subsubsection{N=1}
\label{N1}
For this case the manifold corresponds to a three-dimensional cosmic string space-time. The Green function can be obtained by substituting the Legendre function by its integral representation below \cite{Grad},
\begin{eqnarray}
\label{Q}
Q_{\nu-1/2}(\cosh u)=\frac1{\sqrt{2}}\int_u^\infty dt  \frac{e^{-\nu t}}{\sqrt{\cosh t- \cosh u}} \ ,
\end{eqnarray}
in (\ref{D2c}) and developing the sum over $n$. After some steps we get
\begin{eqnarray}
\label{D3a}
{\cal{D}}_E^{(2)}(x',x)=\frac{e^{i{\bar{N}}(\varphi-\varphi')}}{\alpha(2\pi)^2\sqrt{2rr'}}
\left( 
\begin{array}{cc}
\int_u^\infty  dt  \frac{S^{(+)}(t)}{\sqrt{\cosh t- \cosh u}}&0\\
0&\int_u^\infty dt \frac{S^{(-)}(t)}{\sqrt{\cosh t- \cosh u}}
\end{array} \right)  \ , \nonumber\\ 
\end{eqnarray} 
 where
\begin{eqnarray}
\label{S}
	S^{(\pm)}(t)=\frac{e^{\mp i(\varphi-\varphi')}\sinh(\delta^\pm t/\alpha)-\sinh[(\delta^\pm-1)t/\alpha]}{\cosh(t/\alpha)-\cos(\varphi-\varphi')} \ ,
\end{eqnarray}
being $\delta^\pm=\frac{(1-\alpha)}2\mp\gamma$.

\subsubsection{N=2}
\label{N2}
For the case with $N=2$ the bulk corresponds to a four-dimensional cosmic string space-time. For this case the Green function is very well known. It takes a simple form due to the fact that the respective associated Legendre functions be expressed by
\begin{eqnarray}
	Q_{\nu-1/2}^{1/2}(\cosh u)=i{\sqrt\frac\pi2}\frac{e^{-\nu u}}{{\sqrt{\sinh u}}} \ .
\end{eqnarray} 
Substituting this function into (\ref{D2c}) it is possible to develop the sum on the angular quantum number $n$, providing
\begin{eqnarray}
\label{D3b}
{\cal{D}}_E^{(2)}(x',x)=\frac{e^{i{\bar{N}(\varphi-\varphi')}}}{8\pi^2\alpha rr'\sinh u}
\left( 
\begin{array}{cc}
S^{(+)}(u)&0\\
0&S^{(-)}(u))
\end{array} \right)  \ , \nonumber\\ 
\end{eqnarray}  
with $S^{(\pm)}(u)$ having the same functional form as $S^{(\pm)}(t)$ in (\ref{S}). 

\subsubsection{N=3}
\label{N3}
This case is a new one, it corresponds to a five dimensional cosmic string space-time having a magnetic flux running along the string. For this case the bispinor depends on the associated Legendre function $Q_\nu^1(z)$, which can be expressed in a integral representation by using the relation $Q_\nu^1(z)=(z^2-1)^{1/2}\frac{dQ_\nu(z)}{dz}$ and (\ref{Q}). After some intermediates steps we obtain
\begin{eqnarray}
\label{D3c}
{\cal{D}}_E^{(2)}(x',x)=-\frac{e^{i{\bar{N}(\varphi-\varphi')}}}{{\sqrt{2}}(2\pi)^3\alpha (rr')^{3/2}}\frac{d}{dz}
\left(\begin{array}{cc}
\int_{{\rm arccosh} z}^\infty  dt  \frac{S^{(+)}(t)}{\sqrt{\cosh t-z}}&0\\
0&\int_{{\rm arccosh} z}^\infty dt \frac{S^{(-)}(t)}{\sqrt{\cosh t-z}}
\end{array} \right)|_{z=\cosh u} \ . \nonumber\\
\end{eqnarray}  

\subsubsection{N=4}
\label{N4}
This case is also a new one, it corresponds to a six-dimensional cosmic string space-time. The complete Green function is expressed by a $8\times 8$ diagonal matrix. As in (\ref{N2}), this function can also be written in terms of elementary functions. The reason is because for $N=2$, ${\cal{D}}_E^{(2)}(x',x)$ depends on $Q_{\nu-1/2}^{3/2}(z)$, which is given by
\begin{eqnarray}
	Q_{\nu-1/2}^{3/2}(\cosh u)=-i\sqrt{\frac\pi{2\sinh u}}e^{-\nu u}\frac{(\cosh u +\nu\sinh u)}{\sinh u} \ .
\end{eqnarray}
Substituting this expression into (\ref{D2c}), and after some intermediate steps we arrive to:
\begin{eqnarray}
\label{D3d}
	{\cal{D}}_E^{(2)}(x',x)&=&\frac{e^{i{\bar{N}(\varphi-\varphi')}}}{16\pi^3\alpha(rr')^2\sinh^3u}\times\nonumber\\
&&	\left(\begin{array}{cc}
\cosh u S^{(+)}(u)-\sinh u S'^{(+)}(u)&0\\
0&\cosh u S^{(-)}(u)-\sinh S'^{(-)}(u)
\end{array} \right) \ , \nonumber\\ 
\end{eqnarray}
where the prime denotes derivative with respect to $u$.

\subsection{Special case where $\alpha=\frac1q$ being $q$ an integer number}
\label{Heat0}
In this subsection we shall provide a closed expression for the bispinor considering a non-vanishing mass term for the fermionic field. This will happen for a very special situation when $\alpha$ is equal to the inverse of an integer number. Smith, Davies and Sahni, and Souradeep and Sahni, cited in \cite{scalar}, have shown that when this occur, i.e., for $\alpha=\frac1q$, the scalar Green functions can be given as a sum of $q$ images of the Minkowsiki space-time functions. Recently the image method was also used in \cite{Mello} to provide closed expressions for massive scalar Green functions for a higher-dimensional cosmic string space-time. The mathematical reason for the use of image method in these applications resides in the fact that the order of the Bessel functions which appear in the derivations of the Green functions, become an integer number. Unfortunately, for the fermionic case, the order of the Bessel function depends, besides the integer angular quantum number $n$, also on the the factor $\frac{(1-\alpha)}2$. However, if we consider a charged fermionic field in the presence of a magnetic flux running along the string, an additional effective extra factor will be present, the fractional part of the ratio of the magnetic flux by the quantum one, $\gamma$. For the case where $\gamma$ is equal to $\frac{(1-\alpha)}2$, and $\alpha=1/q$, the order of the Bessel function becomes an integer number, and, in this case it is possible to use the image method to obtain the fermionic Green function in a closed form. Although being a very special situation, the analysis of vacuum polarization effects in this circumstance may shed light on the qualitative behavior of these quantities for non-integer $q$. 

\subsubsection{Heat kernel}
\label{Heat}

Let us now investigate the effective $2\times 2$ diagonal matrix heat kernel, given in the integrand of (\ref{D2}). Using (\ref{D2b}), we have
\begin{eqnarray}
	{\cal{K}}(x,x';s)=\frac{e^{-\frac{(\Delta x)^2+r^2+r'^2}{4s}-M^2s}}{\alpha(4\pi s)^{N/2+1}}\sum_n e^{in\Delta\varphi}
	\left(\begin{array}{cc}
I_{|\nu^+|/\alpha}(rr'/2s)&0\\
0&I_{|\nu^-|/\alpha}(rr'/2s)
\end{array} \right) \ .
\end{eqnarray}

Defining
\begin{eqnarray}
\label{Ka}
 {\bar{K}}^{(+)}(x,x';s)=\sum_n e^{in\Delta\varphi}I_{|\nu^+|/\alpha}(rr'/2s)
 \end{eqnarray}
with $\nu^+=n+\frac{(1-\alpha)}2-{\bar{N}}-\gamma$, we can see that for $\gamma=\frac{(1-\alpha)}2$ and assuming $\alpha=\frac1q$, the order of the modified Bessel function becomes an integer number if $q$ is also an integer number. In this way (\ref{Ka}) can be expressed by \cite{Pru}:\footnote{See the proof in Appendix \ref{app}}
\begin{eqnarray}
\label{Kaa}
 {\bar{K}}^{(+)}(x,x';s)=e^{i{\bar{N}\Delta\varphi}}\sum_n e^{in\Delta\varphi}I_{nq}(rr'/2s)=e^{i{\bar{N}\Delta\varphi}}\ \frac1q\sum_{k=0}^{q-1} e^{\frac{rr'}{2s} \cos\left(\frac{\Delta\varphi}q+\frac{2\pi k}q\right)} \ .
\end{eqnarray}
Analogously, defining 
\begin{eqnarray}
\label{Kb}
 {\bar{K}}^{(-)}(x,x';s)=\sum_n e^{in\Delta\varphi}I_{|\nu^-|/\alpha}(rr'/2s)
 \end{eqnarray}
with $\nu^-=n-\frac{(1-\alpha)}2-{\bar{N}}-\gamma=n-{\bar{N}}-1+\alpha$, we obtain
\begin{eqnarray}
 {\bar{K}}^{(-)}(x,x';s)=e^{i({\bar{N}}+1)\Delta\varphi}\sum_n e^{in\Delta\varphi}I_{nq+1}(rr'/2s) \ .
 \end{eqnarray}
Using the recurrence relation, $I_{\nu+1}(z)=I'_\nu(z)-\frac\nu zI_\nu(z)$ and Eq. (\ref{Kaa}), we get, after some steps, the following expression:
\begin{eqnarray}
\label{Kab}
 {\bar{K}}^{(-)}(x,x';s)=e^{i({\bar{N}}+1-1/q)\Delta\varphi}\ \frac1q\sum_{k=0}^{q-1} e^{\frac{rr'}{2s} \cos\left(\frac{\Delta\varphi}q+\frac{2\pi k}q\right)} e^{-\frac{2ik\pi}{q}} \ .	
\end{eqnarray}

The Green function is given by
\begin{eqnarray}
		{\cal{D}}_E^{(2)}(x',x)=\int_0^\infty ds \ {\cal{K}}(x,x';s) \ .
\end{eqnarray}
Using (\ref{Kaa}), (\ref{Kab}) and with help of \cite{Grad}, this function becomes
\begin{eqnarray}
\label{Dq}
		{\cal{D}}_E^{(2)}(x',x)=\frac{e^{i{\bar{N}}\Delta\varphi}}{(2\pi)^{N/2+1}}M^{N/2}\sum_{k=0}^{q-1}\frac1{(\rho_k)^{N/2}}K_{N/2}(M\rho_k)	\left(\begin{array}{cc}
1&0\\
0&e^{i(1-1/q)\Delta\varphi}e^{-2i\pi k/q}
\end{array} \right) \ ,
\end{eqnarray}
with 
\begin{eqnarray}
\label{rho}
	\rho^2_k=(\Delta x)^2+r^2+r'^2-2rr'\cos\left(\Delta\varphi/q+2\pi k/q\right) \ ,
\end{eqnarray}
being $K_\mu$ the modified Bessel function. 

As we can see, the above Green function of the square of the Dirac operator corresponds, up to a phase factor, to a sum of $q$ images of Minkowski space-time functions. For  $q=1$, and $N=2$, the above expression reduces to the standard four-dimensional Green function
\begin{eqnarray}
{\cal{D}}_E^{(2)}(x',x)=\frac{e^{i{\bar{N}}\Delta\varphi}}{4\pi^2}	\frac M{\rho_0} K_1(M\rho_0)\ I_{(2)} \ .
\end{eqnarray}

We want to finish this section by saying that having obtained the complete Green functions associated with the square of the Dirac operator, for massless and massive fields, the fermionic propagators can be given by applying the Dirac operator on this bispinor according to (\ref{Sf}).

\section{Vacuum expectation value of the energy-momentum tensor}
\label{sec3}
In this section we shall calculate in a explicit form the renormalized vacuum expectation value of the energy-momentum tensor, $\langle T^A_B\rangle_{Ren.}$. As we have already mentioned in the last section, two physical situations will be considered here. Because the metric tensor does not depend of any dimensional parameter, for the massless fermionic field case we can infer that the VEV of the energy-momentum tensor will depend only on the radial distance $r$.  For massive fermionic case, it is expected a dependence on the mass of the field too.
 
The renormalized VEV of the energy-momentum tensor must obey the conservation condition,
	\begin{eqnarray}
\label{CC}
	\nabla_A\langle T^A_B \rangle_{Ren.}=0 \ ,
\end{eqnarray}
and for massless fields provide the correct trace anomaly for space-time of even dimension \cite{Chr}:
\begin{eqnarray}
\label{Tr}
\langle T^A_A \rangle_{Ren.}=\frac1{(4\pi)^{\frac {1+d}2}}{\rm Tr} \ a_{\frac{1+d}2}(x,x) \ .
\end{eqnarray}
However, because this space-time is locally flat and there is no magnetic field in the region outside the string, the coefficients $a_2$ and $a_3$ vanish \cite{Gilkey,JP}. \footnote{For massive fields the trace of the VEV of energy-momentum tensor does not vanish. In fact it is $\langle T^A_A \rangle_{Ren.}=M^2{\rm Tr} \ {\cal{D}}_{Ren.}(x,x)$.} As we shall see, taking into account these informations, it is possible to express all components of the energy-momentum tensor in terms of only one. Here we shall calculate the energy density only.

Using the point-splitting procedure \cite{Birrel}, the VEV of the energy-momentum tensor has the following form:
\begin{eqnarray}
\label{EM}
	\langle T_{AB}(x)\rangle=\frac 14\lim_{x'\to x}{ \rm Tr}\left[{\Gamma}_A(D_B-{\bar D}_{B'})+{\Gamma}_B(D_A-{\bar D}_{A'})\right]S_F(x,x') \ ,
\end{eqnarray}
where $D_M=\nabla_M-ieA_M$, and the bar denotes complex conjugate. Because the dependence of the fermionic Green function on the time variables, the zero-zero component of the energy-momentum tensor reads:
\begin{eqnarray}
\langle T_{00}(x)\rangle=\lim_{x'\to x}{\rm Tr} \ \Gamma_0\partial_0 S_F(x,x') \ , 
\end{eqnarray}
which can be expressed by 
\begin{eqnarray}
\label{T00}
	\langle T_{00}(x)\rangle=-i\lim_{x'\to x}\partial_t^2 \ {\rm Tr}\ {\cal{D}}_F(x',x)=-\lim_{x'\to x}\partial_\tau^2 \ {\rm Tr} \ {\cal{D}}_E(x,x') \ .	
\end{eqnarray}
In the obtainment of the above expression we have to use the fact that the bispinor ${\cal{D}}_F(x',x)$ is diagonal and ${\rm Tr} \ \Gamma^0\Gamma^i {\cal{D}}_E(x,x')=0$. 

Now after these brief comments about general properties of the VEV of the energy-momentum tensor, we shall start explicit calculations of this quantity for massless and massive fermionic fields specifying the dimension of the manifold. 

\subsection{Massless case}
In this subsection we shall calculate the zero-zero component of VEV o the energy-momentum tensor for massless fermionic fields, specifying the dimension of the manifold. Let us start with a three dimensional cosmic string space-time.

\subsubsection{N=1}
\label{N1a}
The vacuum polarization effects associated with massive scalar fields in a three-dimensional conical space-time has bee developed by Guimar\~aes and Linet cited in \cite{scalar}. Because the calculation of the VEV of the energy-momentum tensor involves complicated integrals, the authors give its expression only for the massless case, which coincides with with the result found by Souradeep and Sahni, also cited in \cite{scalar}. However, as far as we know, no calculations regarding to fermionic fields in the presence of a point-like magnetic field have developed in this space-time. 

For this three-dimensional cosmic string space-time the Dirac matrices reads:
\begin{eqnarray}
	\gamma^0=\sigma^3 \ , \ \gamma^1=i{\hat{r}}\cdot{\vec{\sigma}} \ {\rm and} \ \gamma^2=\frac{i}{\alpha r}{\hat{\varphi}}\cdot{\vec{\sigma}} \ ,
\end{eqnarray}
being $\sigma^k$ the Pauli matrices. In this space-time the Euclidean bispinor, ${\cal{D}}_E(x',x)$, is the $2\times 2$ diagonal matrix given below
\begin{eqnarray}
{\cal{D}}_E(x',x)=\frac{e^{i{\bar{N}}(\varphi-\varphi')}}{\alpha(2\pi)^2\sqrt{2rr'}}
\left( 
\begin{array}{cc}
\int_u^\infty  dt  \frac{S^{(+)}(t)}{\sqrt{\cosh t- \cosh u}}&0\\
0&\int_u^\infty dt \frac{S^{(-)}(t)}{\sqrt{\cosh t- \cosh u}}
\end{array} \right)  \ ,
\end{eqnarray} 
where $\cosh u=\frac{(\Delta\tau)^2+r^2+r'^2}{2rr'}$. Let us take first $r'=r$, $\Delta\varphi=0$ in $S^{\pm)}(t)$ given by (\ref{S}), and define new function,
\begin{eqnarray}
	I^{(\pm)}(u)=\int_u^\infty  dt  \frac{S^{(\pm)}(t)}{\sqrt{\cosh t- \cosh u}} \ .
\end{eqnarray}
Introducing a new variable $t:=2 \ {\rm arcsinh(y/\sqrt{2})}$, the above functions is written by
\begin{eqnarray}
	I^{(\pm)}(z)=2\int_{\sqrt{z-1}}^\infty \frac{dy}{\sqrt{2+y^2}}\frac{S^{(\pm)}(2 \ {\rm arcsinh(y/\sqrt{2})})}{\sqrt{y^2-(z-1)}} \ ,
\end{eqnarray}
being $z=\cosh u=1+\frac{(\Delta\tau)^2}{2r^2}$. 

In order to calculate the VEV of the zero-zero component of the energy-momentum tensor, let us write the function $I^{(\pm)}(z)$ as:
\begin{eqnarray}
	I^{(\pm)}(z)&=&I_1^{(\pm)}(z)+I_2^{(\pm)}(z)=2\int_{\sqrt{z-1}}^1 \frac{dy}{\sqrt{2+y^2}}\frac{{\tilde{S}}^{(\pm)}(y)}{\sqrt{y^2-(z-1)}}\nonumber\\
	&+&2\int_1^\infty \frac{dy}{\sqrt{2+y^2}}\frac{{\tilde{S}}^{(\pm)}(y)}{\sqrt{y^2-(z-1)}} \ ,
\end{eqnarray}
where ${\tilde{S}}^{(\pm)}(y)=S^{(\pm)}(t(y))$. Subtracting and adding into the integrand of $I_1^{(\pm)}$ the two first terms of the power expansion 
\begin{eqnarray}
	\frac{{\tilde{S}}^{(\pm)}(y)}{\sqrt{2+y^2}}\approx \frac\alpha y+\frac{[(1-\alpha^2)+6\delta^\pm(\delta^\pm-1)]}{6\alpha}y+ \ ...
\end{eqnarray}
we get
\begin{eqnarray}
	I_1^{(\pm)}=I_1^{(\pm)fin}+I_1^{(\pm)sing} \ ,
\end{eqnarray}
where
\begin{eqnarray}
I_1^{(\pm)fin}(z)=2\int_{\sqrt{z-1}}^1\frac{dy}{\sqrt{y^2-(z-1)}}\left[\frac{{\tilde{S}}^{(\pm)}(y)}{\sqrt{2+y^2}}-\frac\alpha y -\frac{[(1-\alpha^2)+6\delta^\pm(\delta^\pm-1)]}{6\alpha}y\right]
\end{eqnarray}
and
\begin{eqnarray}
	I_1^{(\pm)sing}(z)=2\int_{\sqrt{z-1}}^1\frac{dy}{\sqrt{y^2-(z-1)}}\left[\frac\alpha y +\frac{[(1-\alpha^2)+6\delta^\pm(\delta^\pm-1)]}{6\alpha}y\right] \ .
\end{eqnarray}

As we shall show, $I_1^{(\pm)fin}$ together with $I_2^{(\pm)}$, provide finite contributions to the VEV. The only divergent contribution is contained in $I_1^{(\pm)sing}$, which has the form
\begin{eqnarray}
	I_1^{(\pm)sing}(z)=\frac{2\alpha\beta}{\sqrt{z-1}}+\sqrt{2-z}\frac{[(1-\alpha^2)+6\delta^\pm(\delta^\pm-1)]}{3\alpha}  \ ,
\end{eqnarray}
being
\begin{eqnarray}
	\beta={\rm arctg}\left(\sqrt{\frac{2-z}{z-1}}\right) \ {\rm and} \ z=1+\frac{(\Delta\tau)^2}{2r^2} \ .
\end{eqnarray}
Expanding $	I_1^{(\pm)sing}$ in powers of $\Delta\tau$, and keeping only terms that survive after take the second Euclidean time derivative and coincidence limit, we have
\begin{eqnarray}
\label{Isig}
		I_1^{(\pm)sing}\approx\frac{\alpha\pi\sqrt{2}r}{\Delta\tau}+2(A^\pm-\alpha)-\frac{(3A^\pm+\alpha)}{6r^2}(\Delta\tau)^2+O(\Delta\tau^4) \ ,
\end{eqnarray}
where we use the short notation $A^\pm=\frac{[(1-\alpha^2)+6\delta^\pm(\delta^\pm-1)]}{6\alpha}$.

The first term on the right hand side of (\ref{Isig}) provides
\begin{eqnarray}
	{\cal{D}}_E^{(0)}(x',x)=\frac1{4\pi\Delta\tau}I_{(2)} \ ,
\end{eqnarray}
which coincides with the Euclidean bispinor for a flat three-dimensional space-time. 

The VEV of the zero-zero component of the energy-momentum tensor given by (\ref{T00}) is divergent. In order to obtain a finite and well defined result, we should extract its divergent part. This can be done in a manifest form by subtracting form the complete Green function the usual Euclidean Green function as shown below:
\begin{eqnarray}
	\langle T_0^0(x)\rangle_{Ren.}=\lim_{x'\to x}\partial_\tau^2 \ {\rm Tr} \ \left[{\cal{D}}_E(x,x')-{\cal{D}}_E^{(0)}(x,x')\right]= \lim_{x'\to x}\partial_\tau^2 \ {\rm Tr} \ {\cal{D}}_{Ren.}(x,x') \ .	
\end{eqnarray}

So in order to obtain the VEV above we need to develop three different calculations:
\begin{itemize}
\item The coincidence limit of the Euclidean time derivative of $I_1^{(\pm)ren}(z)=I_1^{(\pm)sing}(z) - \frac{\alpha\pi\sqrt{2}r}{\Delta\tau}$:
\begin{eqnarray}
	\lim_{\Delta\tau\to 0}\partial_\tau^2I_1^{(\pm)ren}(z)=-\frac{(3A^\pm+\alpha)}{3r^2} \ .
\end{eqnarray}
\item The same calculation as above but now for $I_1^{(\pm)fin}(z)$. In the Appendix \ref{app1} we explicitly show that
\begin{eqnarray}
\label{I3}
	\lim_{\Delta\tau\to 0}\partial_\tau^2 I_1^{(\pm)fin}(z)=\frac1{r^2}\int_0^1dy\frac{f^{(\pm)}(y)}{y^3} \ ,
\end{eqnarray}
being
\begin{eqnarray}
\label{f}
	f^{(\pm)}(y)=\left[\frac{{\tilde{S}}^{(\pm)}(y)}{\sqrt{2+y^2}}-\frac\alpha y-A^{(\pm)}y\right] \ .
\end{eqnarray}
\item The last calculation is for $I_2^{(\pm)}(z)$. It is easy to show that
\begin{eqnarray}
		\lim_{\Delta\tau\to 0}\partial_\tau^2 I_2^{(\pm)}(z)=\frac1{r^2}\int_1^\infty\frac{dy}{y^3}\frac{{\tilde{S}}^{(\pm)}(y)}{\sqrt{2+y^2}} \ .
\end{eqnarray}
\end{itemize}

\begin{figure}[tbph]
\begin{center}
\begin{tabular}{cc}
\epsfig{figure=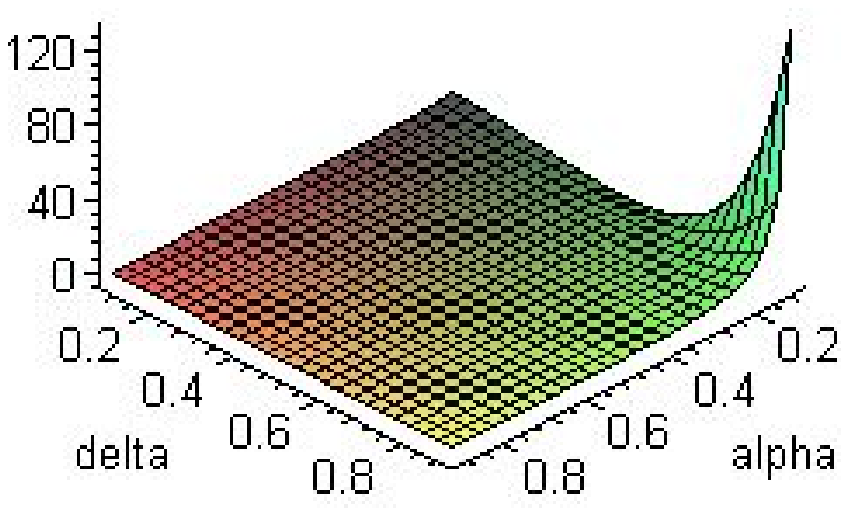, width=7.5cm, height=7.5cm,angle=0} & \quad 
\epsfig{figure=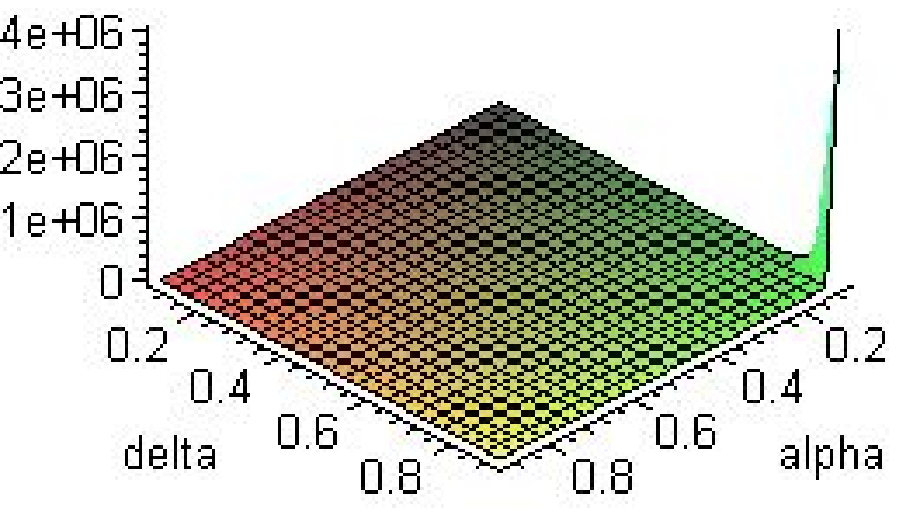, width=7.5cm, height=7.5cm,angle=0}
\end{tabular}
\end{center}
\caption{These graphs provide the behaviors of both integrals in (\ref{T001}) as functions of $\alpha$ and $\gamma$, admitting these parameters varying in the interval $[0.1, \ 0.9]$. The left panel corresponds the first integral and the right panel corresponds to the second one.}
\label{Fig1}
\end{figure}

Finally the renormalized VEV of the zero-zero component of the energy momentum tensor can be written in a simpler form, after substituting the expression for $A^\pm$, using the definition for the function $f^{(\pm)}(y)$ above and expressing $2\alpha/3=2\alpha\int_1^\infty dy/y^4$:
\begin{eqnarray}
\label{T001}
\langle T^0_0(x)\rangle_{Ren.}=\frac1{4\pi^2\alpha\sqrt{2}r^3}\left[\frac{1-\alpha^2-12\gamma^2}{6\alpha}+\int_0^1\frac{dy}{y^3}F_1(y)+ \int_1^\infty\frac{dy}{y^3}F_2(y)\right] \ ,
\end{eqnarray}
where
\begin{eqnarray}
	F_1(y)&=&\left[\frac{{\tilde{S}}^{(+)}(y)+{\tilde{S}}^{(-)}(y)}{\sqrt{2+y^2}}	-\frac{2\alpha}y+\frac{1-\alpha^2-12\gamma^2} {6\alpha}y\right]\nonumber\\
	F_2(y)&=&\left[\frac{{\tilde{S}}^{(+)}(y)+{\tilde{S}}^{(-)}(y)} {\sqrt{2+y^2}}-\frac{2\alpha}y\right] \ .
\end{eqnarray}
The behaviors of the integrals above as functions of $\alpha$ and $\gamma$ are exhibited in figure \ref{Fig1}. 

Having found $F_0^0$ the other components of the renormalized VEV of the energy-momentum tensor can be expressed in terms of it. Assuming 
\begin{eqnarray}
	\langle T^A_B\rangle_{Ren.}=\frac{F^A_B}{r^3} \ ,
\end{eqnarray}
by the diagonal form of the metric tensor, we expect that $F^A_B$ be a diagonal matrix; moreover, the conservation conditions, $\nabla_A\langle T^A_r\rangle_{Ren.}=0$, provides $F_\varphi^\varphi=-2F_r^r$ and $\langle T^A_A\rangle_{Ren.}=0$, $F_0^0=F_r^r$. So we conclude
\begin{eqnarray}
	F^A_B=F_0^0{\rm diag}(1, \ 1, \ -2) \ .
\end{eqnarray}

\subsubsection{N=2} 
This case corresponds to four-dimensional one. For this manifold the renormalized VEV of the energy-momentum tensor associated with fermionic fields, has been analysed by Linet, cited in \cite{ferm}, and by us in \cite{Spin}. So we shall not repeat this calculation; however for completeness we write down the zero-zero component of this tensor:
\begin{equation}
\label{t0}
\left\langle T_0^0\left( x\right)\right\rangle_{ren}=\frac1{2880\pi^{2}\alpha^4 r^4}\left[(\alpha^2- 1)(17\alpha^2+7)+ 120\gamma^2(\alpha^2-2\gamma^2+1)\right] \ .
\end{equation}
Also, having obtained $F_0^0$ the other components of the renormalized VEV of the energy-momentum tensor can be given in terms of it. Assuming a diagonal form for this tensor, a cylindrical symmetry and invariance under a boost along the $x$ direction, the conservation condition $\nabla_A\langle T^A_r\rangle_{Ren.}=0$, provides $F_\varphi^\varphi=-3F_r^r$, and $\langle T^A_A\rangle_{Ren.}=0$, $F^0_0=F^r_r=F_x^x$. So we conclude
\begin{eqnarray}
	F^A_B=F^0_0{\rm diag}(1, \ 1, \ 1, \ -3) \ .
\end{eqnarray}

\subsubsection{N=3}
In braneworld scenario this manifold corresponds to a conical $3-$brane having an one-dimensional extra space transverse to it.\footnote{The five-dimensional cosmic string space-time has been considered in the braneworld theory with metastable graviton on the brane in \cite{Lue}.} The analysis of the VEV of the energy-momentum tensor associated with scalar field in a higher-dimensional cosmic string space-time and in the presence of a magnetic flux has been developed by Guimar\~aes and Linet in \cite{charged}. There, integral representations for massive scalar Green functions have been obtained. Specifically for the four dimensional space, the authors calculated the renormalized vacuum expectation value of the energy-momentum tensor, $\langle T^\nu_\mu\rangle_{Ren.}$, and show that considering the mass of the field equal to zero more suitable expressions are obtained. Moreover, Linet cited in \cite{ferm}, using previous result calculated the VEV of the energy-momentum tensor associated with fermionic fields, relating the parameter $\gamma$, the fractional part of the ratio $\Phi/\Phi_0$, with the conicity parameter $\alpha$. Although a general formalism has been developed, he only provides the vacuum energy density for a massless four-dimensional fermionic field.

The Dirac matrices for a five-dimensional cosmic string space-time, reads
\begin{eqnarray}
\Gamma^0=\gamma^{(0)}, \ \Gamma^r={\hat{r}}\cdot{\vec{\gamma}} \ , \ \Gamma^\varphi=\frac1{\alpha r}{\hat{\varphi}}\cdot{\vec{\gamma}} \ , \ \Gamma^x=\gamma^{(3)} {\rm and}   \ \Gamma^y=i\gamma_5 \ .
\end{eqnarray}
Moreover, the Green function associated with this space is the $4\times 4$ diagonal matrix below
\begin{eqnarray}
\label{D5a}
{\cal{D}}_E(x',x)=
\left( 
\begin{array}{cc}
{\cal{D}}_E^{(2)}(x',x)&0\\
0&{\cal{D}}_E^{(2)}(x',x)
\end{array} \right)  \ ,
\end{eqnarray} 
being
\begin{eqnarray}
{\cal{D}}_E^{(2)}(x',x)=-\frac{e^{i{\bar{N}(\varphi-\varphi')}}}{{\sqrt{2}}(2\pi)^3\alpha (rr')^{3/2}}\frac{d}{dz}
\left(\begin{array}{cc}
I^{(+)}(z)&0\\
0&I^{(-)}(z)
\end{array} \right)|_{z=\cosh u} \ ,
\end{eqnarray}   
with
\begin{eqnarray}
	I^{(\pm)}(z)=\int_{{\rm arccosh} z}^\infty  dt  \frac{S^{(\pm)}(t)}{\sqrt{\cosh t-z}}=2\int_{\sqrt{z-1}}^\infty \frac{dy}{\sqrt{2+y^2}}\frac{{\tilde{S}}^{(\pm)}(y)}{\sqrt{y^2-(z-1)}} \ .
\end{eqnarray}
In the above expression we have introduced the variable $t:=2 \ {\rm arcsinh(y/\sqrt{2})}$ in the second term of the right hand side. As in the analysis developed in subsection \ref{N1a}, we shall divide the integral interval form $[\sqrt{z-1}, \ 1]$ to $[1, \ \infty]$, and define respective integrals by $I_1^{(\pm)}(z)$ and $I_2^{(\pm)}(z)$. Subtracting and adding into the integrand of $I_1^{(\pm)}(z)$ the three first terms of the expansion
\begin{eqnarray}
	\frac{{\tilde{S}}^{(\pm)}(y)}{\sqrt{2+y^2}}\approx \frac\alpha y+A^{(\pm)}y+B^{(\pm)}y^3+ \ ... \ ,
\end{eqnarray}
where $A^\pm=\frac{[(1-\alpha^2)+6\delta^\pm(\delta^\pm-1)]}{6\alpha}$ and $B^{(\pm)}=\frac{(\alpha^2-1)(11\alpha^2+1)+30(\delta^\pm)^2 ((\delta^\pm)^2-1)^2+60\alpha^2\delta^\pm(1-\delta^\pm)}{180\alpha^3}$, we have:
\begin{eqnarray}
	I_1^{(\pm)}(z)&=&2\int_{\sqrt{z-1}}^1 \frac{dy}{\sqrt{y^2-(z-1)}}\left[\frac{{\tilde{S}}^{(\pm)}(y)}{\sqrt{2+y^2}}- \frac\alpha y-A^{(\pm)}y-B^{(\pm)}y^3\right] \nonumber\\
&+&2\int_{\sqrt{z-1}}^1 \frac{dy}{\sqrt{y^2-(z-1)}}\left[\frac\alpha y+A^{(\pm)}y+B^{(\pm)}y^3\right]\nonumber\\
&=&I_1^{(\pm)fin}(z)+I_1^{(\pm)sing}(z)	\ .
\end{eqnarray}

The singular contribution to the VEV is contained only in $I_1^{(\pm)sing}(z)$, which has the form
\begin{eqnarray}
	I_1^{(\pm)sing}(z)=\frac{2\alpha\beta}{\sqrt{z-1}}+2\sqrt{2-z}\left[A^{(\pm)}+B^{(\pm)}\frac{(2z-1)}3\right] \ ,
\end{eqnarray}
being $\beta={\rm arctg}\left(\sqrt{\frac{2-z}{z-1}}\right)$. 

Taking the derivative of $I_1^{(\pm)sing}(z)$ with respect to $z$ and substituting $z=\cosh u=1+\frac{(\Delta x)^2}{2r^2}$,\footnote{In this development we have taken first $r'=r$ and $\varphi'=\varphi$.} we expand the result in powers of $\Delta x$ which survive after taking the second derivative and the coincidence limit. Doing this procedure we get
\begin{eqnarray}
\label{I5}
	\frac{d I_1^{(\pm)sing}(z)}{dz}|_{z=1+\frac{(\Delta x)^2}{2r^2}}&=&-\frac{\alpha\pi r^3\sqrt{2}}{(\Delta x)^3}+\left(B^{(\pm)}-\frac\alpha3-A^{(\pm)}\right)\nonumber\\
&-&\frac{(3\alpha+15B^{(\pm)}+5A^{(\pm)})}{20r^2}(\Delta x)^2+O(\Delta x)^4 \ .
\end{eqnarray}

The other  two contributions for the Green function are given by:
\begin{itemize}
\item The derivative of $I_1^{(\pm)fin}(z)$ which reads
\begin{eqnarray}
	\frac{d I_1^{(\pm)fin}(z)}{dz}=2\frac d{dz}\int_{\sqrt{z-1}}^1\frac{dy}{\sqrt{y^2-(z-1)}}{\tilde{f}}^{(\pm)}(y) \ ,
\end{eqnarray}
where
\begin{eqnarray}
{\tilde{f}}^{(\pm)}(y)=\frac{{\tilde{S}}^{(\pm)}(y)}{\sqrt{2+y^2}}-\frac\alpha y-A^{(\pm)}y-B^{(\pm)}y^3 \ .
\end{eqnarray}
As mentioned in Appendix \ref{app1}, to take the derivative above it is easier if we introduce a new variable $b=\sqrt{z-1}$ into $I_1^{(\pm)fin}(z)$ and change $y\to by$. Doing this procedure we obtain
\begin{eqnarray}
\label{I11}
	\frac{d I_1^{(\pm)fin}(z)}{dz}=\int_b^1\frac{dy}{\sqrt{y^2-b^2}}\frac d{dy}\left(\frac{{\tilde{f}}^{(\pm)}(y)}{y}\right) -\frac{{\tilde{f}}^{(\pm)}(1)}{\sqrt{1-b^2}} \ .
\end{eqnarray}
\item The third contribution comes from the derivative of $I_2^{(\pm)}$, which easily provides
\begin{eqnarray}
\label{I12}
	\frac{d I_2^{(\pm)}(z)}{dz}=\int_1^\infty\frac{dy}{(y^2-b^2)^{3/2}}\frac{{\tilde{S}}^{(\pm)}(y)}{\sqrt{2+y^2}} \ .
\end{eqnarray}
\end{itemize}
The integrals (\ref{I11}) and (\ref{I12}) have to be evaluated at the point $b=\frac{\Delta x}{r\sqrt{2}}$.

Now taking into account all these results the Green function can be expressed by
\begin{eqnarray}
	{\cal{D}}_E(x',x)=-\frac1{\sqrt{2}\alpha(2\pi)^3r^3}
\left( 
\begin{array}{cccc}
K^{(+)}(\Delta x)&0&0&0\\
0&K^{(-)}(\Delta x)&0&0\\
0&0&K^{(+)}(\Delta x)&0\\
0&0&0&K^{(-)}(\Delta x)
\end{array} \right)  \ ,
\end{eqnarray}
where
\begin{eqnarray}
	K^{(\pm)}(z)=\frac{d I_1^{(\pm)sing}(z)}{dz}+\frac{d I_1^{(\pm)fin}(z)}{dz}+\frac{d I_2^{(\pm)}(z)}{dz} \ .
\end{eqnarray}

The formal expression for the VEV of the zero-zero component of the energy-momentum tensor is given by (\ref{T00}). However this expression is divergent, and the term responsible for that is precisely the first term on the right hand side of (\ref{I5}), which when substituting into into (\ref{D5a}) provides
\begin{eqnarray}
	{\cal{D}}_E^{(0)}(x',x)=\frac1{8\pi^2(\Delta x)^3}I_{(4)} \ .
\end{eqnarray}
So, in order to obtain a finite and well defined result to this VEV we should extract its divergent part. Here, this also can be done in a manifest form by subtracting form the complete Green function the usual Euclidean Green given above:
\begin{eqnarray}
	\langle T_0^0(x)\rangle_{Ren.}=\lim_{x'\to x}\partial_\tau^2 \ {\rm Tr} \ \left[{\cal{D}}_E(x,x')-{\cal{D}}_E^{(0)}(x,x')\right]= \lim_{x'\to x}\partial_\tau^2 \ {\rm Tr} \ {\cal{D}}_{Ren.}(x,x') \ .	
\end{eqnarray}

\begin{figure}[tbph]
\begin{center}
\begin{tabular}{cc}
\epsfig{figure=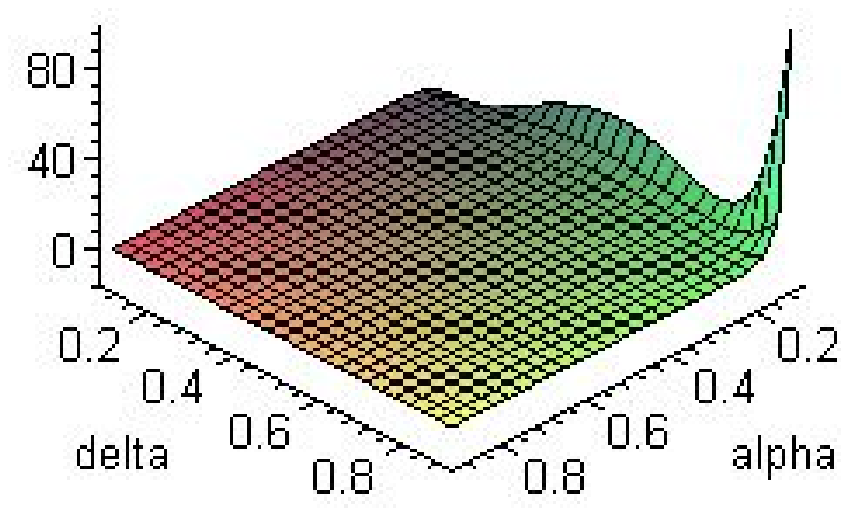, width=7.5cm, height=7.5cm,angle=0} & \quad 
\epsfig{figure=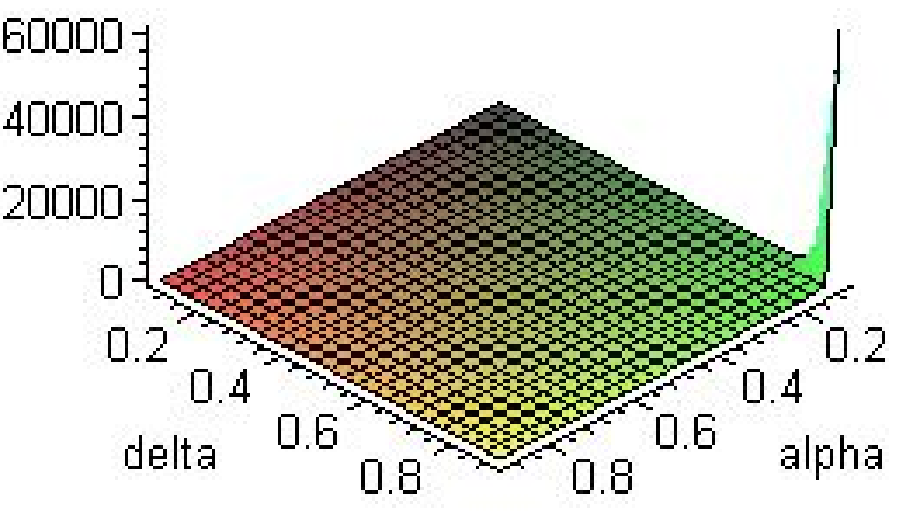, width=7.5cm, height=7.5cm,angle=0}
\end{tabular}
\end{center}
\caption{These graphs exhibit the behaviors of the integrals in (\ref{T003}) as functions of $\alpha$ and $\gamma$, admitting these parameters in the interval $[0.1, \ 0.9]$. The left panel corresponds the first integral and the right panel corresponds to the second one.}
\label{Fig2}
\end{figure}

Finally we are in position to write down the the zero-zero component of the renormalized VEV of the energy-momentum tensor in a simpler form. Substituting the expressions for $A^{(\pm)}$ and $B^{(\pm)}$, using definition for the function $\tilde{f}^{(\pm)}(y)$ and some intermediates steps we get:
\begin{eqnarray}
\label{T003}
	\langle T^0_0(x)\rangle_{Ren.}&=&\frac1{8\pi^3\alpha\sqrt{2}r^5}\left[\frac{(\alpha^2-1)(23\alpha^2-7)-120\gamma^2(1-3\alpha^2-2\gamma^2)} {240\alpha^2} \right.\nonumber\\
	&-&\left.3\int_0^1\frac{dy}{y^5}F_3(y)-3\int_1^\infty\frac{dy}{y^5}F_4(y)\right] \ ,
\end{eqnarray}
where
\begin{eqnarray}
	F_3(y)&=&\frac{{\tilde{S}}^{(+)}(y)+{\tilde{S}}^{(-)}(y)}{\sqrt{2+y^2}}-\frac{2\alpha}y+\frac{(1-\alpha^2-12\gamma^2)}{6\alpha}y\nonumber\\
	&+&\frac{[(\alpha^2-1)(17\alpha^2+7)+120\gamma^2(1+\alpha^2-2\gamma^2)]}{720\alpha^3}y^3 \nonumber\\
	F_4(y)&=&\left[\frac{{\tilde{S}}^{(+)}(y)+{\tilde{S}}^{(-)}(y)} {\sqrt{2+y^2}}-\frac{2\alpha}y\right] \ .
\end{eqnarray}
Here also it is not possible to provide analytical expressions for the integrals above. The numerical behaviors for both integrals are plotted in figure \ref{Fig2} as functions of the parameters $\alpha$ and $\gamma$.

The other components of the energy-momentum tensor can be obtained from the zero-zero one, by using the conservation condition (\ref{CC}), the absence of trace anomaly (\ref{Tr}) and the boosts symmetry in the directions parallel to the string. Writing 
\begin{eqnarray}
	\langle T^A_B(x)\rangle_{Ren.}=\frac{F^A_B}{r^5} \ ,
\end{eqnarray}
we can easily show that
\begin{eqnarray}
	F^A_B=F^0_0{\rm diag}(1, \ 1, \ -4, \ 1, \ 1) \ .
\end{eqnarray}

\subsubsection{N=4}
As we have mentioned beforee \cite{Cohen,Ruth}, the gravitational effects associated with strings have been considered as responsible for compactification from six to four dimensional space-time. In this scenario the bulk is represented by a conical $3-$brane transverse to a two dimensional space.

The manifold analyzed by us here presents a simpler geometrical structure. The bulk corresponds to an idealized conical $3-$brane transverse to a flat two dimensional space. 

The explicit analysis of the VEV of the energy-momentum tensor developed here is a new one. Fortunately, it becomes much simpler when compared with the previous analysis because the respective Green function can be expressed in terms of elementary functions. 

In this six-dimensional cosmic string space-time, the  Dirac matrices are in the form (\ref{gamma}), and the Green function is expressed in terms of a diagonal $8\times 8$ matrix:
\begin{eqnarray}
\label{D6a}
{\cal{D}}_E(x',x)=
\left( 
\begin{array}{cccc}
{\cal{D}}_E^{(2)}(x',x)&0&0&0\\
0&{\cal{D}}_E^{(2)}(x',x)&0&0\\
0&0&{\cal{D}}_E^{(2)}(x',x)&0\\
0&0&0&{\cal{D}}_E^{(2)}(x',x)
\end{array} \right)  \ ,
\end{eqnarray} 
being
\begin{eqnarray}
{\cal{D}}_E^{(2)}(x',x)&=&\frac{e^{i{\bar{N}(\varphi-\varphi')}}}{16\pi^3\alpha(rr')^2\sinh^3u}\times\nonumber\\
&&	\left(\begin{array}{cc}
\cosh u S^{(+)}(u)-\sinh u S'^{(+)}(u)&0\\
0&\cosh u S^{(-)}(u)-\sinh S'^{(-)}(u)
\end{array} \right) \ , \nonumber\\ 
\end{eqnarray}   
with $S^{(\pm)}(u)$ having the same functional form as $S^{(\pm)}(t)$ given in $(\ref{S})$.

As we have already discussed, the calculation of the renormalized VEV of the zero-zero component of the energy-momentum tensor can be calculated by applying the second order Euclidean time derivative on the renormalized Green function:
\begin{eqnarray}
	\langle T_0^0(x)\rangle_{Ren.}=\lim_{x'\to x}\partial_\tau^2 \ {\rm Tr} \ \left[{\cal{D}}_E(x,x')-{\cal{D}}_E^{(0)}(x,x')\right]= \lim_{x'\to x}\partial_\tau^2 \ {\rm Tr} \ {\cal{D}}_{Ren.}(x,x') \ ,	
\end{eqnarray}
where ${\cal{D}}_E^{(0)}(x,x')$ is promptly obtained from (\ref{D6a}) by taking $\alpha=1$ and $\gamma=0$. After some intermediate steps we obtain for the energy density the result below:
\begin{eqnarray}
\label{t00}
\langle T_0^0(x)\rangle_{ren}&=-&\frac1{120960\pi^{3}\alpha^6  r^6}\left[-367(1-\alpha^2)^3+12(189\gamma^2+76)(1-\alpha^2)^2 \right. \nonumber \\
 && \left. +144(35\gamma^4-49\gamma^2-4)(1-\alpha^2)+1344\gamma^2(\gamma^2-1)(\gamma^2-4) \right] \ .
\end{eqnarray}

The other components of the energy-momentum tensor can be computed by using the conservation condition, the absence of trace anomaly and the boost symmetry along the directions parallel to the string. So we have:
\begin{eqnarray}
	\langle T^A_B(x)\rangle_{Ren.}=\frac{F^A_B}{r^6} \ ,
\end{eqnarray}
where
\begin{eqnarray}
	F^A_B=F^0_0{\rm diag}(1, \ 1, \ -5, \ 1, \ 1, \ 1) \ 
\end{eqnarray}
being $F^0_0$ given in (\ref{t00}).

\subsection{Massive case}
\label{MC}
Here we shall analyze the VEV associated with massive field under the circumstance specified in \ref{Heat0}. In order to that we shall adopt the bispinor (\ref{Dq}) to calculate the fermionic Green function. However, before to start the calculation let us analyse this bispinor in the coincidence limit. Taking $x'\to x$ we verify that the result is divergent and that the divergence comes exclusively form the $k=0$ component. So, in order to obtain a finite and well defined result we should apply some renormalization prescription. This procedure can be applied in a manifest form by subtracting from (\ref{Dq}) its $k=0$ component. So the renormalized Green function is given by
\begin{eqnarray}
\label{Dren}
		{\cal{D}}_{Ren.}^{(2)}(x',x)=\frac{e^{i{\bar{N}}\Delta\varphi}}{(2\pi)^{N/2+1}}M^{N/2}\sum_{k=1}^{q-1}\frac1{(\rho_k)^{N/2}}K_{N/2}(M\rho_k)	\left(\begin{array}{cc}
1&0\\
0&e^{i(1-1/q)\Delta\varphi}e^{-2i\pi k/q}
\end{array} \right) \ .
\end{eqnarray}

Before to start the calculation of the renormalized VEV of the energy-momentum tensor, we shall analyse the behavior of the renormalized bispinor in the coincidence limit.

Considering first the upper diagonal component of (\ref{Dren}), we have
\begin{eqnarray}
{\cal{D}}_{Ren.}^{(+)}(x,x)=\frac{M^{N/2}}{(2\pi)^{N/2+1}(2r)^{N/2}}\sum_{k=1}^{q-1}\frac{K_{N/2}(2Mr\sin(k\pi/q))}{(\sin(k\pi/q))^{N/2}} \ .	
\end{eqnarray}
We can see that the ${\cal{D}}_{Ren.}^{(+)}(x,x)$ is positive everywhere. Now we would like to analyse this expression in two limiting cases:
\begin{itemize}
\item In the limit $Mr>>1/\sin(k\pi/q)$, the main contribution comes from $k=1$ and $k=q-1$ and the leading term is 
\begin{eqnarray}
	{\cal{D}}_{Ren.}^{(+)}(x,x)\approx\frac{M^{\frac{N-1}{2}}}{(4\pi)^{\frac{N+1}{2}}}\frac{e^{-2Mr\sin(\pi/q)}}{(r\sin(\pi/q))^{\frac{N+1}{2}}} \ .
\end{eqnarray}
\item For the massless limit, 
\begin{eqnarray}
\label{D11}
	{\cal{D}}_{Ren.}^{(+)}(x,x)=\frac{\Gamma(N/2)}{(4\pi)^{N/2+1}r^N}\sum_{k=1}^{q-1}\frac1{\sin(k\pi/q))^N} \ .
\end{eqnarray}
\end{itemize}

For even $N$, the summation on the right hand side of (\ref{D11}) can be developed by using the formulas \cite{Mello} 
\begin{eqnarray}
\label{Ir}
	I_{N+2}(x)=\frac{I''_N(x)+N^2I_N(x)}{N(N+1)} \ {\rm and} \ I_2(x)=\frac{q^2}{\sin^2(qx)}-\frac1{\sin^2(x)} \ .
\end{eqnarray}
for the sum
\begin{eqnarray}
	I_N(x)=\sum_{k=1}^{q-1}\frac1{\sin^N(x+k\pi/q)} \ .
\end{eqnarray}
In particular for a four dimensional space, $N=2$, $I_2(0)=\frac{(q^2-1)}3$.  So we have
\begin{eqnarray}
\label{D.1}
		{\cal{D}}_{Ren.}^{(+)}(x,x)=\frac{q^2-1}{48\pi^2r^2} \ .
\end{eqnarray}
For a six-dimensional space, $I_4(x)$ is a long expression obtained by the recurrence relation above, however we find $I_4(0)=\frac{(q^2-1)(q^2+11)}{45}$, consequently
\begin{eqnarray}
\label{D.2}
		{\cal{D}}_{Ren.}^{(+)}(x,x)=\frac{(q^2-1)(q^2+11)}{2880\pi^3r^4} \ .
\end{eqnarray}
The above results, Eq.s (\ref{D.1}) and (\ref{D.2}), are analytical functions of $q$, and by analytical continuation are valid for all arbitrary values of $q$.

A similar analysis can also developed for the lower diagonal component of (\ref{Dren}). So let consider
\begin{eqnarray}
{\cal{D}}_{Ren.}^{(-)}(x,x)=\frac{M^{N/2}}{(2\pi)^{N/2+1}(2r)^{N/2}}\sum_{k=1}^{q-1}\frac{K_{N/2}(2Mr\sin(k\pi/q))} {(\sin(k\pi/q))^{N/2}} e^{-2ik\pi/q} \ .	
\end{eqnarray}
In the two limiting cases above we have
\begin{itemize}
\item For $Mr>>1/\sin(k\pi/q)$ we find
\begin{eqnarray}	{\cal{D}}_{Ren.}^{(-)}(x,x)\approx\frac{M^{\frac{N-1}{2}}}{(4\pi)^{\frac{N+1}{2}}}\frac{\cos(2\pi/q)e^{-2Mr\sin(\pi/q)}} {(r\sin(\pi/q))^{\frac{N+1}{2}}} \ .
\end{eqnarray}
\item For the massless limit, 
\begin{eqnarray}
\label{D22}
	{\cal{D}}_{Ren.}^{(-)}(x,x)=\frac{\Gamma(N/2)}{(4\pi)^{N/2+1}r^N}\sum_{k=1}^{q-1}\frac{e^{-2ik\pi/q}}{\sin(k\pi/q))^N} \ .
\end{eqnarray}
\end{itemize}
In order to develop the sum in (\ref{D22}) for $N$ even, the relevant formula is
\begin{eqnarray}
	K_N=\sum_{k=1}^{q-1}\frac{e^{-2ik\pi/q}}{\sin(k\pi/q))^N}=\sum_{k=1}^{q-1}\frac{\cos(2k\pi/q)}{\sin(k\pi/q))^N}=I_N(0)-2I_{N-2}(0) \ .
\end{eqnarray}
For a four-dimensional space $K_2=I_2(0)-2I_0(0)=\frac{(q^2-1)}3-2(q-1)=\frac{(q-1)(q-5)}{3}$, consequently
\begin{eqnarray}
		{\cal{D}}_{Ren.}^{(-)}(x,x)=\frac{(q-1)(q-5)}{48\pi^2r^2} \ .
\end{eqnarray}
For six-dimensional space, $K_4=I_4(0)-2I_2(0)=\frac{(q^2-1)(q^2-19)}{45}$ and
\begin{eqnarray}
		{\cal{D}}_{Ren.}^{(-)}(x,x)=\frac{(q^2-1)(q^2-19)}{2880\pi^3r^4} \ .
\end{eqnarray}

Now let us proceed the calculation of $\langle T^M_N(x)\rangle_{Ren.}$. Although we shall concentrate the analysis of this quantity for a six-dimensional cosmic string spacetime, below we present its zero-zero component for a general value of $N$. By using previous result,
\begin{eqnarray}
\label{T0m}
	\langle T_0^0(x)\rangle_{Ren.}=\lim_{x'\to x}\partial_\tau^2 \ {\rm Tr} \ {\cal{D}}_{Ren.}(x,x') 	\ ,
\end{eqnarray}
and substituting the Euclidean bispinor expressed by its diagonal form given in terms of (\ref{Dren}), after some intermediate steps we obtain 
\begin{eqnarray}
\label{T1m}
\langle T_0^0(x)\rangle_{Ren}=-\frac{2M^{N/2+1}}{\pi(4\pi)^{N/2}r^{N/2+1}}\sum_{k=1}^{q-1}\frac{\cos^2(k\pi/q)}{(\sin(k\pi/q) )^{N/2+1}}K_{N/2+1}(2Mr\sin(\pi k/q)) \ .
\end{eqnarray}
From (\ref{T1m}) we can observe that the energy density is non-positive quantity everywhere, vanishing only for $q=2$. For this case the contributions from the upper and lower component of the bispinor (\ref{Dren}) cancel each other. Moreover in the limit $Mr>>1/\sin(k\pi/q)$ the leading order is
\begin{eqnarray}
\langle T_0^0(x)\rangle_{Ren}\approx-\frac1{\pi(4\pi)^{(N-1)/2}\ r}\left(\frac Mr\right)^{(N+1)/2}\times\frac{\cos^2(\pi/q)}{(\sin(\pi/q))^{(N+3)/2}} e^{-2Mr\sin(\pi/q)}	\ ,
\end{eqnarray}
with exponentially suppressed behavior.

In the massless limit case, the energy density became expressed in terms of elementary functions,
\begin{eqnarray}
	\langle T_0^0(x)\rangle_{Ren}=-\frac{\Gamma(N/2+1)}{\pi(4\pi)^{N/2}r^{N+2}}\sum_{k=1}^{q-1}\frac{\cos^2(k\pi/q)}{(\sin(k\pi/q) )^{N+2}} \ .
\end{eqnarray}
For even $N$, the summation on the right side of the above equation can be expressed in terms of analytical functions of $q$ as shown below:
 \begin{eqnarray}
	\langle T_0^0(x)\rangle_{Ren}=-\frac{\Gamma(N/2+1)}{\pi(4\pi)^{N/2}r^{N+2}}\left(I_{N+2}(0)-I_N(0)\right) \ .
\end{eqnarray}
For $N=4$, we easily find from the recurrence formula (\ref{Ir}), $I_6(0)=\frac1{945}(q^2-1)(2q^4+23q^2+191)$, consequently
\begin{eqnarray}
\label{t001}
	\langle T_0^0(x)\rangle_{Ren}=-\frac{(q^2-1)(q^4+q^2-20)}{3780\pi^3r^6} \ .
\end{eqnarray}
The above quantity is positive for $1<q \ < 2$. \footnote{Substituting $\alpha=\frac1q$ and $\gamma=\frac{q-1}{2q}$ into (\ref{t00}) we re obtain (\ref{t001}).}

Now going back to the general expression (\ref{T1m}), in figure $3$ are plotted the behaviors of $\langle T_0^0\rangle_{Ren.}/M^{N+2}$ as function of $Mr$ for different values of $N$ and the parameter $q$. As we have discussed, for large value of $Mr$ the energy density tends to zero exponentially. The right panel explicitly shows that the modulus of the energy density increases when $q$ becomes larger.
\begin{figure}[tbph]
\label{figj2}
\begin{center}
\begin{tabular}{cc}
\epsfig{figure=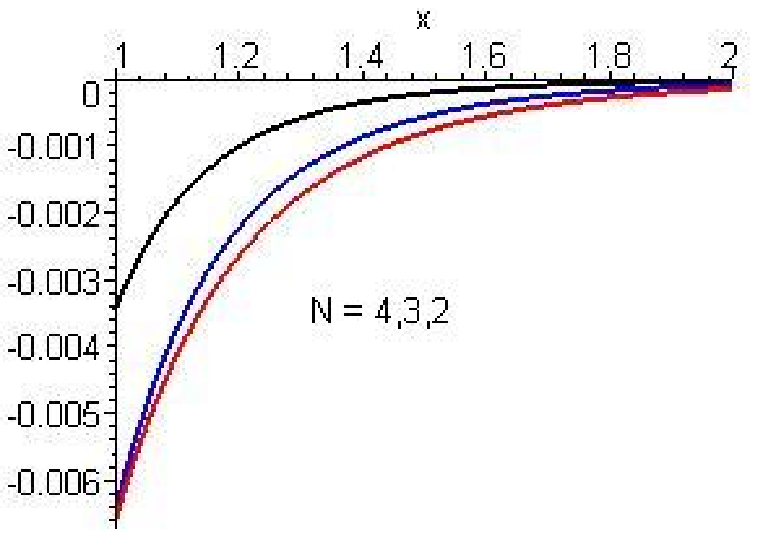, width=7.5cm, height=7.5cm,angle=0} & \quad 
\epsfig{figure=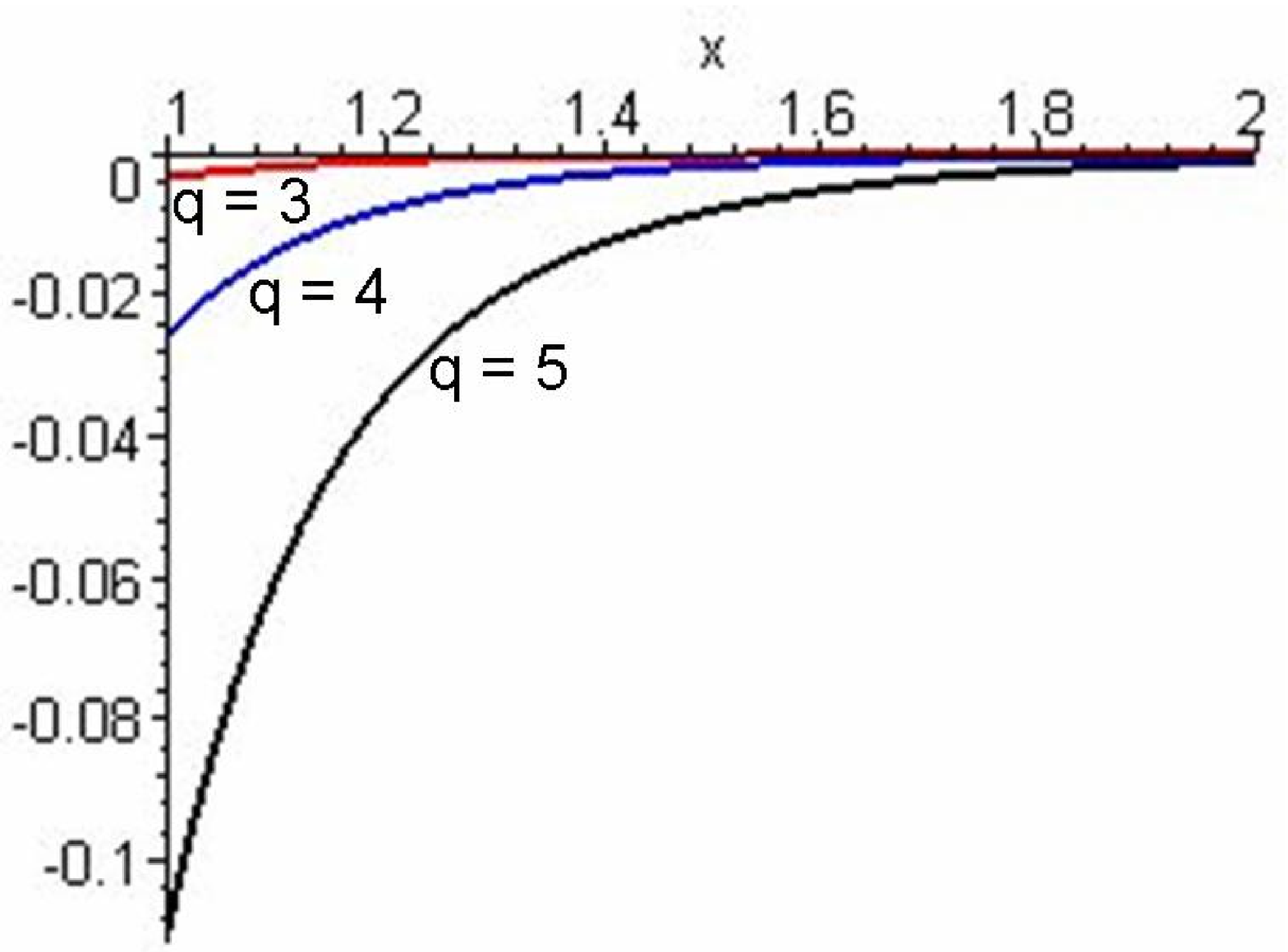, width=7.5cm, height=7.5cm,angle=0}
\end{tabular}
\end{center}
\caption{These graphs represent the dependences of $\langle T_0^0\rangle_{Ren.}/M^{N+2}$ with $Mr$ for different values of the parameters $N$ and $q$. The left panel corresponds to $q=4$ and $N=2, \ 3, \ 4$, and the right panel corresponds to $N=4$ and $q=3, \ 4, \ 5$.}
\end{figure}

Now after this discussion about the behavior of the energy density for a general value of $N$, let us specialize in a six-dimensional cosmic string spacetime. Below we present the most important analytical results: 
\begin{itemize}
\item Taking advantage of previous result we can promptly write down the zero-zero component of the VEV of the energy-momentum tensor. It is
\begin{eqnarray}
\langle T_{0}^{0}(x)\rangle_{Ren.}=-\frac{M^{3}}{8\pi^{3}r^{3}}\sum_{k=1}^{q-1}\frac{\cos^2\left(k\pi/q \right)}{\sin^{3}\left( k\pi/q\right)}K_{3}\left[ 2Mr\sin \left(k\pi/q \right)\right] \ .
\end{eqnarray}
\item The radial pressure, $\langle T^r_r\rangle$, can be calculated as shown below:
\begin{equation}
\langle T_{rr}(x)\rangle_{Ren.}=\frac{1}{2}\lim_{x'\to x}Tr\left[\Gamma_{r}\left(\partial_{r}-\partial_{r'} \right)\right]S_{F}\left( x,x'\right) \ .
\end{equation}
By using (\ref{Sf}) to calculate the  Feynman propagator, we can write
\begin{eqnarray}
\langle T_{rr}(x)\rangle_{Ren.}&=&\frac{1}{2}\lim_{x'\rightarrow x}Tr\left[\left(\partial_{r}-\partial_{r'}\right)\left[\partial_r +\frac{i}{\alpha r}\Sigma_{(8)}^{(3)}\left( \partial_\varphi-ieA_\varphi\right)\right.\right.\nonumber\\
&-&\left.\left.\frac{1}{2\alpha r}\left(1-\alpha \right)\right]\right]
D_{E}\left( x,x'\right)  \ .
\end{eqnarray} 
Using $eA_\varphi={\bar{N}}+\gamma$, $\gamma=\frac{1-\alpha}2$, $\alpha=1/q$ and the identities
\begin{eqnarray}
\Gamma_{r}\Gamma^{r}=-I \ , \quad  \Gamma_{r}\Gamma^{\theta}=-\frac{i}{\alpha r}\Sigma_{(8)}^{(3)} \ {\rm and} \quad \Gamma_{r}\Gamma^{\theta}\Pi_{\theta}=-\frac{i}{2 \alpha r}(1-\alpha)I \ ,
\end{eqnarray}
after a long calculation we find
\begin{eqnarray}
\label{Trr}
\langle {T_{r}^{r}}(x)\rangle_{Ren.}=-\frac{M^{3}}{8\pi^{3}r^{3}}\sum_{k=1}^{q-1}\frac{\cos^2(k\pi/q)}{\sin^{3}(k\pi/q)}K_{3}[2Mr\sin(k\pi/q)] \ .
\end{eqnarray}
\item From the conservation condition $\nabla_A\langle T^A_r\rangle_{Ren.}=0$, the azimuthal pressure, $\langle T^\varphi_\varphi\rangle$, can be obtained in terms of the radial one
\begin{eqnarray}
	\langle T^\varphi_\varphi(x)\rangle_{Ren.}=\partial_r\left(r\langle T^r_r(x)\rangle_{Ren.}\right) \ .
\end{eqnarray}
Substituting (\ref{Trr}) into the above expression we get
\begin{eqnarray}
\langle T_\varphi^\varphi(x)\rangle_{Ren.}&=&\frac{M^3}{8\pi^3r^3}\sum_{k=1}^{q-1}\frac{\cos^2(k\pi/q)}{\sin^3(k\pi/q)}\left\{5K_3[2Mr\sin(k\pi/q)] \right. \nonumber\\
&+&\left.2K_{2}[2Mr\sin(k\pi/q)]Mr\sin(k\pi/q)\right\} \ .
\end{eqnarray}
\item For the pressures along the directions parallel to the string we have (no summation over $i$)
\begin{eqnarray}
	\langle T^i_i\rangle_{Ren.}=\langle T^0_0\rangle_{Ren.} \ , \ i=3, \ 4, \ 5 \ .
\end{eqnarray}
\item Now on basis on the results obtained, we can verify the correct trace of the energy-momentum tensor:
\begin{eqnarray}
	\langle T_A^A(x)\rangle_{Ren.}=\frac{M^{4}}{4\pi^{3}r^{2}}\sum_{k=1}^{q-1}\frac{\cos^2(k\pi/q)}{\sin^2(k\pi/q)}K_{2}\left[ 2Mr\sin(k\pi/q) \right]=M^{2}\hbox{Tr}D_{Ren.}(x,x) \ .
\end{eqnarray}
\end{itemize}

\section{Conclusion and Discussions}
\label{conc}
In this paper we have investigated the local one-loop quantum gravity effects associated with charged fermionic fields in a higher-dimensional cosmic string spacetime. As the first steps in the evaluation of the renormalized VEV of the energy-momentum tensor, in Section \ref{sec2} we calculated the general Euclidean Green function, which is expressed in terms of not-solvable integral for a general situation. However, two limiting cases provide more workable expressions which  were explicitly analyzed: $i)$ the massless case, and $ii)$ the massive one with $\alpha=\frac1q$, being $q$ an integer number, and $\gamma=\frac{1-\alpha}2$. For the latter, we were able to express the Green function in terms of a finite sum of modification Bessel functions, $K_\mu$.

In Section \ref{sec3}, we considered the VEV of the energy-momentum tensor. For the massless case, we calculated, in explicit form, the energy-density for different dimensions. For even dimensions, $\langle T^0_0\rangle_{Ren.}$ is expressed in terms of elementary functions; however for odd dimensions, it is given in terms of integrals expressions. For these situations we provided numerical analysis, as shown in figures $1$ and $2$. Being $q$ an integer number and $\gamma=\frac{q-1}{2q}$, the corresponding renormalized VEV of the energy-momentum tensor is expressed in a closed form for massive fields. To our knowledge no such closed expression has been given before in the literature. For an arbitrary value for $N$, we explicitly shown the behaviors of the energy-density for several limiting cases. For points near the string, it behaves as $1/r^{N+2}$, and for points far away from it, presents an exponentially suppressed behavior. Finally, for a six-dimensional space, we calculated completely all components of the energy-momentum tensor, and shown they provide the trace identity.

An interesting point which deserves to be mentioned is that the effects on the renormalized VEV of the energy-momentum tensor due to the conical structure of the spacetime and the magnetic interaction, may cancel each other. This was explicitly observed in Subsection \ref{MC} for $q=2$ and $\gamma=\frac14$, and can be also verified in the results found for the energy-densities given in (\ref{t0}) and (\ref{t00}). In fact, for these specials values, and taking $\Delta\varphi=0$, the upper and lower components of (\ref{D3b})  can be expressed as
\begin{eqnarray}
	{\cal{D}}_E^{(+)}(x',x)&=&\frac1{4\pi^2r^2}\frac{\cosh u}{\sinh^2 u} \nonumber\\
	{\cal{D}}_E^{(-)}(x',x)&=&\frac1{4\pi^2r^2}\frac1{\sinh^2 u} \ ,
\end{eqnarray}
with $\cosh(u)$ given by (\ref{cs}), which coincides with the respective quantities found by image method,
\begin{eqnarray}
	{\cal{D}}_E^{(+)}(x',x)&=&\frac1{4\pi^2}\left[\frac1{\rho_0^2}+\frac1{\rho_1^2}\right] \nonumber\\
	{\cal{D}}_E^{(-)}(x',x)&=&\frac1{4\pi^2}\left[\frac1{\rho_0^2}-\frac1{\rho_1^2}\right] \ ,
\end{eqnarray}
being $\rho_0$ and $\rho_1$ given in (\ref{rho}). In this situation the contributions on the renormalized VEV of the energy-momentum tensor due to the upper and lower components of the Feynman propagator cancel each other.

The renormalization procedure adopted in this paper to calculate the vacuum expectation value of the fermionic energy-momentum tensor was the point-splitting one. In this sense we had analysed the behavior of this quantity at the coincidence limit and extracted all the divergences in manifest form by using the Hadamard function.\footnote{Because the cosmic string spacetime is locally flat the Hadamard function coincides with the usual Green one.} It has been proved that this procedure provides finite and well defined results \cite{Wald}. An alternative regularization procedure can also be used. The local $\zeta-$function procedure leads essentially to the same conclusions as given by the point-splitting one \cite{Moretti}. The DeWitt-Schwinger expansion of the Green function for an $n-$dimensional curved spacetime is given in \cite{Birrel,Chr}. So, by using the dimensional regularization, it follows from the general expression that for the cosmic string spacetime this expansion contains a single term only,
\begin{eqnarray}
	G_F^{DS}(x',x)=\frac{i\pi}{(4\pi i)^{n/2}}\left(-\frac{z}{2im^2}\right)^{1-n/2}H_{n/2-1}^{(2)}(z) \ ,
\end{eqnarray}
where $H_\nu^{(2)}$ is the Hankel function of the second kind, $z^2=-2m^2\sigma$ being $\sigma$ the one-half of the square of the geodesic distance. The above function coincides with the Minkowski Green function. The subtracted contribution due to this function renormalizes the cosmological constant.

All these regularization procedures lead to the same finite result for renormalized quantities.

\section*{Acknowledgment}

ERBM thanks Conselho Nacional de Desenvolvimento Cient\'\i fico e Tecnol\'ogico (CNPq.)  for partial financial support, FAPESQ-PB/CNPq. (PRONEX) and FAPES-ES/CNPq. (PRONEX).

\appendix
\section{The generalization formula of the generating function of modified Bessel Function $I_\nu$}
\label{app}
The method of images applied in (\ref{Heat}) to calculate the Green function associated with massive fermionic fields in a conical space-time and in the presence of a magnetic flux running along the string, is an application of the generalization formula of the generating function for the modified Bessel functions of integer order. It well known that
\begin{eqnarray}
	e^{x/2(t+1/t)}=\sum_{n=-\infty}^\infty I_n(x)t^n \ .
\end{eqnarray}
Taking $t=e^{i\varphi}$ we have
\begin{eqnarray}
\label{a1}
	e^{x\cos\varphi}=\sum_{n=-\infty}^\infty I_n(x)e^{in\varphi} \ .
\end{eqnarray}
Making the change $\varphi\to \varphi+\frac{2\pi k}{q}$ in (\ref{a1}), and summing over $k$ from $0$ until $q-1$, we may write
\begin{eqnarray}
\sum_{k=0}^{q-1}e^{x\cos(\varphi+2\pi k/q)}=\sum_{n=-\infty}^\infty I_n(x)e^{in\varphi}\sum_{k=0}^{q-1}e^{i2\pi nk/q} \ .	
\end{eqnarray}
Now using the identity 
\begin{eqnarray}
	\sum_{k=0}^{q-1}e^{i2\pi nk/q}=q\sum_{m=-\infty}^\infty\delta_{n,mq}
\end{eqnarray}
we obtain
\begin{eqnarray}
	\sum_{k=0}^{q-1}e^{x\cos(\varphi+2\pi k/q)}=q\sum_{m=-\infty}^\infty I_{mq}(x)e^{imq\varphi} \ .
\end{eqnarray}

\section{Explicit calculation of $F^0_0$}
\label{app1}
In order to find the expression (\ref{I3}) for the limit case analysed, our first steps is to define a new variable $b=\sqrt{z-1}=\frac{\Delta\tau}{\sqrt{2}r}$. In terms of this variable we may express $I_1^{(\pm)fin}(z)$ as:
\begin{eqnarray}
	I_1^{(\pm)fin}(z)=2\int_b^\infty\frac{dy}{\sqrt{y^2-b^2}}f^{(\pm)}(y) \ ,
\end{eqnarray}
being $f^{(\pm)}(y)$ given by (\ref{f}). Using the fact that $\partial_\tau^2=(1/2r^2)\partial_b^2$ we have
\begin{eqnarray}
	\partial_\tau^2I_1^{(\pm)fin}(z)=\frac1{r^2}\frac{d^2}{db^2} \int_b^\infty\frac{dy}{\sqrt{y^2-b^2}}f^{(\pm)}(y) \ .
\end{eqnarray}
Because the integrand is divergent at the point $y=b$, we have to change the variable $y\to by$ before applying the Leibnitz formula for derivative of an integral. So the second derivative has the form below,
\begin{eqnarray}
	\frac{d^2}{db^2} \int_b^\infty\frac{dy}{\sqrt{y^2-b^2}}f^{(\pm)}(y)=\frac1{b^2}\int_b^1\frac{dy \ y^2}{\sqrt{b^2-y^2}}f''^{(\pm)}(y)- \frac{f'^{(\pm)}(1)}{b^2\sqrt{1-b^2}}+\frac{f^{(\pm)}(1)(1-2b^2)}{b^2(1-b^2)^{3/2}} \ ,
\end{eqnarray}
which can be written as
\begin{eqnarray}
		\frac{d^2}{db^2} \int_b^\infty\frac{dy}{\sqrt{y^2-b^2}}f^{(\pm)}(y)&=&\int_b^1\frac{dy}{{\sqrt{y^2-b^2}}}\left(f''^{(\pm)}(y) -\frac{f'^{(\pm)}(y)}{y}+\frac{f^{(\pm)}(y)}{y^2}\right)\nonumber\\
		&-&\frac{f'^{(\pm)}(1)}{\sqrt{1-b^2}}-\frac{f^{(\pm)}(1)b^2}{(1-b^2)^{3/2}} \ .
\end{eqnarray}
Taking the limit $b\to 0$ we obtain
\begin{eqnarray}
	\frac{d^2}{db^2} \int_b^\infty\frac{dy}{\sqrt{y^2-b^2}}f^{(\pm)}(y)\to \int_0^1 dy\frac{f^{(\pm)}(y)}{y^3} \ .
\end{eqnarray}
We can see that the integral above is finite because for small value of $y$, $f^{(\pm)}(y)\to \frac{(\alpha^2-1)(11\alpha^2+1)+30(\delta^\pm)^2 ((\delta^\pm)^2-1)^2+60\alpha^2\delta^\pm(1-\delta^\pm)}{180\alpha^3}y^3$.


\end{document}